\documentclass[prd,aps,preprint,amsmath,nofootinbib,amssymb,eqsecnum,showkeys,tightenlines]{revtex4-1}
\usepackage[colorlinks=true,urlcolor=blue,anchorcolor=blue,citecolor=blue,filecolor=blue,linkcolor=blue,menucolor=blue,pagecolor=blue,linktocpage=true]{hyperref} 
\usepackage{slashed}
\usepackage{epsfig,latexsym,cancel,amssymb,amsmath,verbatim,mathrsfs}
\usepackage{color}
\usepackage{graphicx}

\usepackage[inline]{enumitem}
\usepackage[multidot]{grffile}  
\usepackage{dcolumn}
\usepackage{bm}
\usepackage{amsmath}
\usepackage{amsfonts}
\usepackage{amssymb}
\usepackage{color}
\usepackage{latexsym}
\usepackage{slashed} 
\usepackage{pstricks}
\usepackage{indentfirst}
\usepackage{mathrsfs}
\usepackage{multirow}
\usepackage{epsfig,psfrag}
\usepackage{subfigure}
\usepackage{setspace} 
\usepackage[utf8]{inputenc} 
\usepackage[scientific-notation=true]{siunitx} 

\newcommand{\SUtwoL}{\mathrm{SU}(2)_\mathrm{L}}
\newcommand{\UoneY}{\mathrm{U}(1)_\mathrm{Y}}

\allowdisplaybreaks 

\graphicspath{{fig/}}

\begin{document}

\title{Scalar Quintuplet Minimal Dark Matter \\ with Yukawa Interactions:
Perturbative up to the Planck Scale}
\author{Chengfeng Cai$^1$}
\author{Zhaofeng Kang$^2$}
\author{Zhu Luo$^1$}
\author{Zhao-Huan Yu$^{1,3}$}\email[]{yuzhaoh5@mail.sysu.edu.cn}
\author{Hong-Hao Zhang$^{1,}$}\email[]{zhh98@mail.sysu.edu.cn}
\affiliation{$^1$School of Physics, Sun Yat-Sen University, Guangzhou 510275, China}
\affiliation{$^2$School of physics, Huazhong University of Science and Technology, Wuhan 430074, China}
\affiliation{$^3$ARC Centre of Excellence for Particle Physics at the Terascale,
School of Physics, The~University of Melbourne, Victoria 3010, Australia}

\begin{abstract}
We confront the perturbativity problem in the real scalar quintuplet minimal dark matter model.
In the original model, the quintuplet quartic self-coupling inevitably hits a Landau pole at a scale $\sim10^{14}~\si{GeV}$, far below the Planck scale.
In order to push up this Landau pole scale, we extend the model with a fermionic quintuplet and three fermionic singlets which couple to the scalar quintuplet via Yukawa interactions.
Involving such Yukawa interactions at a scale $\sim 10^{10}~\si{GeV}$ can not only keep all couplings perturbative up to the Planck scale, but can also explain the smallness of neutrino masses via the type-I seesaw mechanism.
Furthermore, we identify the parameter regions favored by the condition that perturbativity and vacuum stability are both maintained up to the Planck scale.
\end{abstract}

\maketitle

\tableofcontents

\clearpage

\section{Introduction}\label{sect1}

One of the biggest mysteries of Nature, dark matter (DM) has drawn much attention from astrophysicists, cosmologists, and particle physicists.
Among various guesses at the identity of the DM particle, the most extensively studied class of DM candidates is weakly interacting massive particles (WIMPs), because they can naturally explain the observed DM relic abundance via the thermal production mechanism in the early Universe~\cite{Bertone:2004pz,Feng:2010gw,Young:2016ala,Arcadi:2017kky}. WIMP models can be easily constructed by introducing a dark sector with electroweak $\SUtwoL$ multiplets. Introducing one nontrivial $\SUtwoL$ multiplet leads to the so-called minimal dark matter (MDM) models~\cite{Cirelli:2005uq,Cirelli:2007xd,Cirelli:2008id,Cirelli:2009uv,Hambye:2009pw,Buckley:2009kv,Cai:2012kt,Earl:2013jsa,Cirelli:2014dsa,Ostdiek:2015aga,Cirelli:2015bda,Garcia-Cely:2015dda,Cai:2015kpa,DelNobile:2015bqo}, which only involve the minimal content of new fields. Introducing more than one $\SUtwoL$ multiplet results in a richer phenomenology, but the models are much more complicated~\cite{Mahbubani:2005pt,DEramo:2007anh,Enberg:2007rp,Cohen:2011ec,Fischer:2013hwa,Cheung:2013dua,Dedes:2014hga,Fedderke:2015txa,Calibbi:2015nha,Freitas:2015hsa,Yaguna:2015mva,Tait:2016qbg,Horiuchi:2016tqw,Banerjee:2016hsk,Cai:2016sjz,Abe:2017glm,Lu:2016dbc,Cai:2017wdu,Maru:2017otg,Liu:2017gfg,Egana-Ugrinovic:2017jib,Xiang:2017yfs,Voigt:2017vfz,Wang:2017sxx}.

The philosophy of the MDM models is to extend the standard model (SM) in a minimal way to involve dark matter~\cite{Cirelli:2005uq}.
For this purpose, a fermionic or scalar $\SUtwoL\times\UoneY$ multiplet in a representation $(\mathbf{n},Y)$ is introduced. The potential DM candidate would be the electrically neutral component that should be the lightest new state.
If the dimension of the $\SUtwoL$ representation $n$ is large enough to forbid dangerous decay operators, this neutral state would be able to play the role of the DM particle without imposing an artificial $Z_2$ symmetry.
In other words, the stability of DM can be protected by an accidental $Z_2$ symmetry due to a large dimension of the $\SUtwoL$ representation and the Lorentz invariance.
As proposed in the original paper~\cite{Cirelli:2005uq}, the minimal choices are a quintuplet ($n=5$) for the fermionic case~\cite{Cirelli:2007xd,Cirelli:2009uv,Cirelli:2015bda,Ostdiek:2015aga,Garcia-Cely:2015dda} and a septuplet ($n=7$) for the scalar case~\cite{Hambye:2009pw,Cai:2012kt,Earl:2013jsa,Garcia-Cely:2015dda,Cai:2015kpa}. Adding new electroweak multiplets will push up the $\SUtwoL$ gauge coupling $g_2$ when it runs to high scales. Nevertheless, the MDM model with a quintuplet fermion or a septuplet scalar can keep $g_2$ perturbative up to $\sim 10^{19}~\si{GeV}$~\cite{Cirelli:2005uq} ($\sim 10^{16}-10^{17}~\si{GeV}$~\cite{DiLuzio:2015oha}) based on one-loop (two-loop) $\beta$ functions.

Scalar MDM models are quite different from fermionic ones, since scalars will bring in more coupling terms. Such complexity has caused the neglect of a dangerous decay operator for the septuplet scalar model in the original consideration: the dimension-5 operator $\Phi^3 H^\dag H$ will induce DM decay at loop level~\cite{DiLuzio:2015oha}. Therefore, the accidental $Z_2$ symmetry in the septuplet scalar model is not totally strict. Nonetheless, one can always introduce an artificial $Z_2$ symmetry to make the model work again, but considering $n=7$ would not be special any more. In this case, discussing a triplet ($n=3$) or quintuplet ($n=5$) real\footnote{The term ``real'' means that the multiplet is self-conjugated. A electroweak multiplet with even $n$ must be complex, and hence allows more interaction terms.} scalar multiplet with $Y=0$ would be more economic.
The real scalar triplet model has been studied thoroughly in Refs.~\cite{Araki:2011hm,Ayazi:2014tha,Khan:2016sxm}, while the real scalar quintuplet model is less explored and will be the main topic of this paper.

In such scalar MDM models, scalar coupling terms may lead to another problem. Solutions to the renormalization group equations (RGEs) show that the scalar self-interaction couplings will go to infinity, \textit{i.e.}, a Landau pole (LP) will show up, at an energy scale far below the Planck scale~\cite{Hamada:2015bra,Cai:2015kpa,Khan:2016sxm}.
With two independent septuplet self-interaction terms, the real scalar septuplet model hits a Landau pole at a scale around $10^8~\si{GeV}$ if the DM particle mass is fixed to satisfy the observed relic abundance.
In our previous work~\cite{Cai:2015kpa}, we attempted to push up the LP scale via introducing Yukawa couplings between the scalar septuplet and extra fermionic multiplets.
A bonus of these extra fermions is to explain the smallness of neutrino masses through the type-III seesaw mechanism~\cite{Foot:1988aq}.
We found that such a setup can push up the LP scale to $\sim 10^{14}$~GeV at best.


On the other hand, a real scalar quintuplet lives in a smaller representation and has only one independent self-interaction term.
Consequently, the quintuplet couplings should evolve slower and reach a Landau pole at a higher scale.
If extra fermionic multiplets are introduced, we may even push the LP scale above the Planck scale.
Besides, such fermions could be used to explain the tiny neutrino masses via the type-I seesaw mechanism~\cite{Minkowski:1977sc,GellMann:1980vs,Yanagida:1979as,Mohapatra:1979ia}.
In this work, we will explore these possibilities in the real scalar quintuplet MDM model.
For completeness, we will also discuss the constraint from the observed DM relic abundance, the bounds from direct and indirect detection experiments, and the stability of the electroweak vacuum.


The paper is organized as follows. In Sec.~\ref{sec:5plet}, we introduce the quintuplet MDM model, and discuss its phenomenological constraints and the LP scale. In Sec.~\ref{sec:551}, we study an extension with extra fermions for pushing up the LP scale, and discuss the constraints from perturbativity and vacuum stability. Conclusions and discussions are given in Sec.~\ref{sect4}. Appendix~\ref{app:beta} gives the $\beta$ functions and initial values of SM couplings, while Appendix~\ref{app:SE} gives the detailed calculation of the Sommerfeld enhancement effect.

\section{Real scalar quintuplet MDM Model}\label{sec:5plet}


\subsection{Model details}

In the real scalar quintuplet MDM model, the dark sector only involves a  real scalar quintuplet $\Phi$ with $Y=0$, which can be expressed as
\begin{eqnarray}
\Phi=\frac{1}{\sqrt{2}}(\Delta^{(2)},\ \Delta^{(1)},\ \Delta^{(0)},\ \Delta^{(-1)},\ \Delta^{(-2)})^\mathrm{T}.\label{phidef}
\end{eqnarray}
The self-conjugate condition implies $(\Delta^{(Q)})^\ast=\Delta^{(-Q)}$.
The real scalar $\Delta^{(0)}$ is a viable DM candidate.
The gauge covariant derivative of $\Phi$ is
\begin{eqnarray}
D_\mu \Phi=\partial_\mu \Phi-ig_2W^a_\mu\tau^a \Phi,\label{gaugecovdef}
\end{eqnarray}
where $\tau^a$ are generators for the  $\SUtwoL$ representation $\mathbf{5}$:
\begin{eqnarray}\setlength{\arraycolsep}{.4em}
\tau^1 &=&
\begin{pmatrix}
 & -1 &  &  &  \\
-1 &  & -\sqrt{6}/2  &  &  \\
 &  -\sqrt{6}/2 &  & \sqrt{6}/2 &  \\
 &  & \sqrt{6}/2 &  & 1 \\
 &  &  & 1 &
\end{pmatrix},\quad
\tau^2 =
\begin{pmatrix}
 & i &  &  &  \\
-i &  & \sqrt{6}i/2  &  &  \\
 &  -\sqrt{6}i/2 &  & -\sqrt{6}i/2 &  \\
 &  & \sqrt{6}i/2 &  & -i \\
 &  &  & i &
\end{pmatrix},
\\
\tau^3 &=& \mathrm{diag}(2,1,0,-1,-2).
\end{eqnarray}
Thus, the covariant kinetic term for $\Phi$ can be expanded as
\begin{eqnarray}\label{gaugecouple}
\mathcal{L}_\mathrm{kin}&=&(D_\mu \Phi)^\dag D^\mu \Phi
\nonumber\\
&=&\frac{1}{2}(\partial_\mu\Delta^{(0)})^2+\sum_{Q=1}^2(\partial_\mu\Delta^{(Q)})(\partial^\mu\Delta^{(-Q)})
+\sum_{Q=1}^2(QeA^\mu+Qg_2c_\mathrm{W} Z^\mu)\Delta^{(-Q)}i\overleftrightarrow{\partial_\mu}\Delta^{(Q)}\nonumber\\
&&-g_2\big[W^{+,\mu}(\sqrt{2}\Delta^{(-2)}i\overleftrightarrow{\partial_\mu}\Delta^{(1)}
+\sqrt{3}\Delta^{(-1)}i\overleftrightarrow{\partial_\mu}\Delta^{(0)})+\mathrm{h.c.}\big]
\nonumber\\
&&+(e^2A_\mu A^\mu+g_2^2c_\mathrm{W}^2 Z_\mu Z^\mu+2eg_2c_\mathrm{W} A_\mu Z^\mu)
\sum_{Q=1}^2Q^2\Delta^{(Q)}\Delta^{(-Q)}
\nonumber\\
&&+g_2^2 W^+_\mu W^{-,\mu}\big[3(\Delta^{(0)})^2+5\Delta^{(1)}\Delta^{(-1)}
+2\Delta^{(2)}\Delta^{(-2)}\big]
\nonumber\\
&&-g_2^2\bigg\{ W^+_\mu(s_\mathrm{W} A^\mu + c_\mathrm{W} Z^\mu)
(\sqrt{3}\Delta^{(0)}\Delta^{(-1)}+3\sqrt{2}\Delta^{(1)}\Delta^{(-2)})
\nonumber\\
&&\qquad\quad + W^+_\mu W^{+,\mu} \bigg[\frac{3}{2}(\Delta^{(-1)})^2
-\sqrt{6}\Delta^{(0)}\Delta^{(-2)}\bigg]
+\mathrm{h.c.}\bigg\},
\end{eqnarray}
where $s_\mathrm{W}\equiv\sin\theta_\mathrm{W}$, $c_\mathrm{W}\equiv\cos\theta_\mathrm{W}$, and  $\overleftrightarrow{\partial_\mu}$ is defined as $F\overleftrightarrow{\partial_\mu} G=F\partial_\mu G-G\partial_\mu F$.

In order to protect the stability of $\Delta^{(0)}$, we require that $\Phi$ is odd under a $Z_2$ symmetry, while all SM fields are even.
The scalar potential is constructed by $\Phi$ and the SM Higgs doublet $H$.
Since the operator $\Phi^\dag\tau^a\Phi$ vanishes due to the self-conjugation condition, the general form of the potential respecting the $Z_2$ symmetry is given by only five independent terms:
\begin{equation}\label{potential}
V=\mu^2 H^\dag H+m^2\Phi^\dag\Phi+\lambda(H^\dag H)^2+\lambda_2(\Phi^\dag\Phi)^2
+\lambda_3(H^\dag H)(\Phi^\dag\Phi).
\end{equation}
Therefore, this model just brings in two couplings, $\lambda_2$ and $\lambda_3$, and one mass parameter $m$ as new free parameters.
We assume that the vacuum expectation value (VEV) of the Higgs field is nonzero, while the VEV of $\Phi$ remains zero.
Then the minimization of the potential implies two conditions, $\mu^2<0$ and $m^2-{\lambda_3\mu^2}/{(2\lambda)}\geq0$.
As in the SM, the VEV of the Higgs doublet is $\langle H\rangle=(0,{v}/{\sqrt{2}})^\mathrm{T}$ with $v\equiv\sqrt{{-\mu^2}/{\lambda}}=246.22~\si{GeV}$.

After the Higgs field acquires a VEV, the $\lambda_3(H^\dag H)(\Phi^\dag\Phi)$ term contributes equally to the masses of all the $\Phi$ components. Therefore, at the tree level all components are degenerate with a shifted mass $m_0$, given by
\begin{eqnarray}
m_{0}^2=m^2+\frac{\lambda_3}{2}v^2.
\end{eqnarray}
Electroweak one-loop corrections break this degeneracy, making $\Delta^{(1)}$ and $\Delta^{(2)}$ slightly heavier than $\Delta^{(0)}$.
When $m_0 \gg m_Z$, the mass difference between $\Delta^{(Q)}$ and $\Delta^{(0)}$ is~\cite{Cirelli:2005uq}
\begin{equation}\label{eq:m_corr}
m_Q - m_0 = Q^2 \Delta m,
\end{equation}
where $\Delta m=\alpha_2 m_W\sin^2({\theta_\mathrm{W}}/{2})\simeq 167~\mathrm{MeV}$, with $\alpha_2\equiv g_2^2/{(4\pi)}$.

Vacuum stability (VS) sets a stringent constraint on the model.
The philosophy is that the potential should remain bounded from below as the couplings evolve to high energies.
The VS conditions can be obtained by means of the copositive criteria~\cite{Kannike:2012pe}:
\begin{equation}\label{VS1}
\lambda\geq0,\quad
\lambda_2\geq0,\quad
\lambda_3+2\sqrt{\lambda\lambda_2}\geq0.
\end{equation}

\subsection{Experimental constraints}

The observation of the DM relic abundance sets a constraint on the $\Delta^{(0)}$ mass $m_0$.
Assuming DM is thermally produced in the early Universe, its relic abundance can be expressed as~\cite{Griest:1990kh}
\begin{eqnarray}
\Omega_\mathrm{DM}h^2\simeq \frac{1.07\times 10^9~\si{GeV^{-1}}}{J(x_\mathrm{F})\sqrt{g_\ast}M_{\mathrm{Pl}}}~~\text{with}~~ J(x_\mathrm{F})=\int^\infty_{x_\mathrm{F}}\frac{\langle\sigma_{\mathrm{eff}}v\rangle}{x^2}dx,
\end{eqnarray}
where $M_{\text{Pl}}$ is the Planck mass,  $x_\mathrm{F}$ is the freeze-out parameter, $g_\ast$ is the total number of effectively relativistic degrees of freedom, and $\left<\sigma_{\mathrm{eff}}v\right>$ the effective thermally averaged annihilation cross section accounting for the coannihilation effect.

For $m_0\gg m_h$, annihilation and coannihilation into gauge and Higgs bosons in the $s$-wave are dominant, leading to the following result~\cite{Hambye:2009pw}:
\begin{eqnarray}\label{eq:sv_eff}
\langle\sigma_{\mathrm{eff}}v\rangle\simeq \frac{66\pi \alpha_2^2}{5m_0^2}+\frac{\lambda_3^2}{80\pi m_0^2}.
\label{sv}
\end{eqnarray}
We take $x_\mathrm{F}\simeq 25$ and $\sqrt{g_\ast}\simeq 10.33$ for $T\sim \mathcal{O}(\si{TeV})$, and calculate the prediction to the relic abundance in the quintuplet MDM model.

A more proper treatment is to consider the Sommerfeld enhancement (SE) effect for DM freeze-out, following the strategy in Refs.~\cite{Cirelli:2007xd,Cirelli:2015bda}. Annihilation and coannihilation channels in the dark sector are categorized by the electric charges of the two-body states.
Enhancement factors can be computed in various categories via numerically solving the Schr\"{o}dinger equations for the two-body states. The inclusion of the SE effect would increase the effective annihilation cross section, and hence reduce the relic abundance for fixed model parameters.
Details of the calculation are summarized in Appendix~\ref{app:SE}.

\begin{figure}[!t]
\centering\includegraphics[width=.6\textwidth]{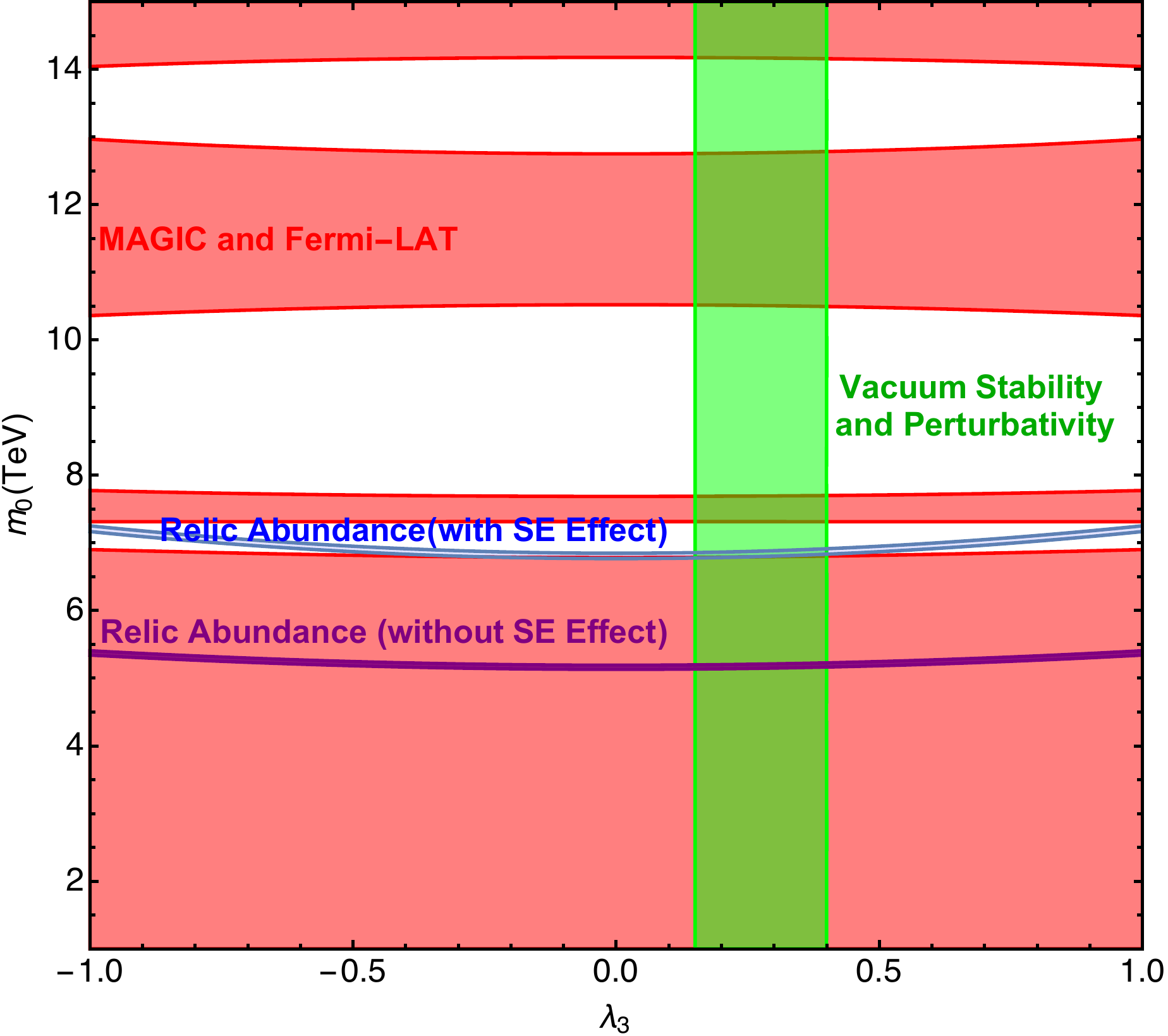}
\caption{Constraints on the real scalar quintuplet MDM model in the $\lambda_3$-$m_0$ plane.
The purple (blue) strip corresponds to the 1$\sigma$ range of the DM relic abundance measured by the Planck experiment~\cite{Ade:2015xua} without (with) the Sommerfeld enhancement effect.
The red regions are excluded at $95\%$ CL by the combined result of the MAGIC and Fermi-LAT indirect detection experiments~\cite{Ahnen:2016qkx}.
The green band indicates the region satisfying the vacuum stability and
perturbativity conditions in the 5-5-1 model described in Sec.~\ref{sec:551}.}
\label{fig:constr}
\end{figure}

Assuming thermally produced $\Delta^{(0)}$ particles in the early Universe fulfill the observed relic abundance, $\Omega_\mathrm{DM}h^2=0.1193\pm 0.0014$~\cite{Ade:2015xua}, the $\Delta^{(0)}$ mass $m_0$ can be constrained within a narrow range, depending on the coupling $\lambda_3$. 
As shown in Fig.~\ref{fig:constr}, a slightly larger $m_0$ is favored for a larger $|\lambda_3|$.
If the SE effect is not taken into account,
the favored $m_0$ is $\sim 5~\si{TeV}$.
After including the SE effect, the favored value\footnote{This value may be slightly modified if the bound state formation effect is also considered~\cite{Mitridate:2017izz}.} is $m_0\sim 7~\si{TeV}$.

Other constraints come from DM direct and indirect detection experiments.
Direct detection uses the response to DM-nucleon scattering.
The only tree-level diagram of the spin-independent $\Delta^{(0)}$-nucleon scattering process is mediated by the Higgs boson, but its cross section is suppressed by $m_0^2$ and thus insignificant unless $\lambda_3$ is very large.
On the other hand, according to the discussions in Ref.~\cite{Hisano:2015rsa}, the gauge loops lead to a DM-nucleon cross section of $\sim 2\times10^{-46}~\si{cm^2}$ for a fermionic quintuplet.
As the gauge interactions of the scalar quintuplet are similar to those of the fermionic one, we may expect that the cross section for the scalar case would also be around this value\footnote{In order to give an accurate DM-nucleon cross section, a detailed calculation for loop diagrams is needed. But such a calculation would be beyond the scope of this paper. We will leave it to a further study.}.
Current direct detection experiments, such as PandaX-II~\cite{Cui:2017nnn} and XENON1T~\cite{Aprile:2018dbl}, have set an upper bound of $\sim 5\times10^{-45}~\si{cm^2}$ on the DM-nucleon cross section for a DM particle mass of $7~\si{TeV}$.
Thus, the scalar quintuplet model can evade current direct searches, but should be well tested in near future experiments.

For indirect detection of DM annihilation in space, the dominant process is $\Delta^{(0)}\Delta^{(0)}\rightarrow W^+W^-$.
The SE effect on such a process is even more significant, since the velocities of Galactic DM particles are much lower than those at the freeze-out epoch. 
Details of the related calculation are also given in Appendix~\ref{app:SE}.
Using the 95\% CL exclusion limit on annihilation cross section obtained by the combined MAGIC and Fermi-LAT $\gamma$-ray observations of dwarf satellite galaxies assuming DM totally annihilating into $W^+W^-$~\cite{Ahnen:2016qkx}, we find that indirect detection experiments have given very stringent constraints, as shown by the red regions in Fig.~\ref{fig:constr}.
Nonetheless, three windows around $m_0 \simeq 7~\si{TeV}$, $8-10~\si{TeV}$, and $13-14~\si{TeV}$ survive.
As a result, the parameter regions suggested by the observation of relic abundance are still available.

\subsection{$\beta$ functions and the Landau pole scale}

RGE evolution of couplings are determined by $\beta$ functions.
In Appendix~\ref{app:beta}, we list the $\beta$ functions in the SM, as well as initial values of SM couplings.
When the renormalization scale $\mu$ goes above a scale $\Lambda_\mathrm{s} \sim m_0$, the effect of the real scalar quintuplet should be involved.
Hereafter we will adopt $\Lambda_\mathrm{s} = 10~\si{TeV}$, as  suggested by the observed relic abundance.
At one-loop level, the real scalar quintuplet MDM model introduces new terms in the $\beta$ functions of the $\SUtwoL$ gauge coupling $g_2$ and the Higgs quartic coupling $\lambda$, while the $\beta$ functions of the other SM couplings do not change.
Here we list the $\beta$ functions that are relevant in the following analysis:~\cite{Hamada:2015bra}
\begin{eqnarray}
\beta_{g_1}&=&\beta_{g_1}^\mathrm{SM},\quad
\beta_{g_2}=\beta_{g_2}^\mathrm{SM}+\frac{1}{16\pi^2}\frac{5}{3}g_2^3,\quad
\beta_{g_3}=\beta_{g_3}^\mathrm{SM},\quad
\beta_{y_t}=\beta_{y_t}^\mathrm{SM},
\label{betafunc_1}\\
\beta_{\lambda}&=&\beta_{\lambda}^\mathrm{SM}+\frac{1}{16\pi^2}\frac{5}{2}\lambda_3^2,\quad
\beta_{\lambda_2}=\frac{1}{16\pi^2}[26\lambda_2^2+108g_2^4-72g_2^2\lambda_2+2\lambda_3^2],
\label{betafunc_2}\\
\beta_{\lambda_3}&=&\frac{1}{16\pi^2}\left[12\lambda\lambda_3+14\lambda_2\lambda_3+4\lambda_3^2+18g_2^4-\lambda_3\left(\frac{81}{2}g_2^2+\frac{9}{10}g_1^2-6y_t^2\right)\right].\label{betafunc_3}
\end{eqnarray}
Note that $g_1$ is related to the $\UoneY$ gauge coupling $g'$ by $g_1\equiv \sqrt{5/3}g'$.
The solution to $\alpha_2 = g_2^2/(4\pi)$ for $\mu > \Lambda_\mathrm{s}$ is just
\begin{equation}
\alpha_2^{-1}(\mu) = \alpha_2^{-1}(m_Z)
-\frac{b_2^\mathrm{SM}}{2\pi}\ln\frac{\Lambda_\mathrm{s}}{m_Z}
-\frac{b_2^\mathrm{s}}{2\pi}\ln\frac{\mu}{\Lambda_\mathrm{s}},
\end{equation}
where $b_2^\mathrm{SM}=-19/6$ and $b_2^\mathrm{s}=b_2^\mathrm{SM}+5/3=-3/2$ are coefficients corresponding to $\beta_{g_2}^\mathrm{SM}$ in Eq.~\eqref{SM:betafunc_1} and $\beta_{g_2}$ in Eq.~\eqref{betafunc_1}, respectively.

Below we analyze the Landau pole scale.
The large coefficients of the $\lambda_2^2$ and $g_2^4$ terms in the beta function of $\lambda_2$ drive $\lambda_2$ to divergence  at high energy scales.
At one-loop level, $g_1$ and $y_t$ do not enter the beta function of $\lambda_2$, while $\lambda_2$ and $\lambda_3$ do not contribute to the beta functions of $g_1$ and $y_t$.
Therefore, $g_1$ and $y_t$ remain small at high energy scales.
For an approximate analysis, we can safely neglect $g_1$ and $y_t$ in Eqs.~\eqref{betafunc_2} and \eqref{betafunc_3}.
Thus, the RGEs for $\lambda$, $\lambda_2$, and $\lambda_3$ become
\begin{eqnarray}
\frac{d\lambda}{dt'}&=&24\lambda^2+\frac{9}{8}g_2^4-9g_2^2\lambda+\frac{5}{2}\lambda_3^2,\\
\frac{d\lambda_2}{dt'}&=&26\lambda_2^2+108g_2^4-72g_2^2\lambda_2+2\lambda_3^2,\\
\frac{d\lambda_3}{dt'}&=&4\lambda_3^2+18g_2^4-\frac{81}{2}g_2^2\lambda_3+12\lambda\lambda_3+14\lambda_2\lambda_3,
\end{eqnarray}
where $t'\equiv (4\pi)^{-2} \ln(\mu/\Lambda_\mathrm{s})$.
For convenience, we define three functions $f_i$ by $\lambda=f_1g_2^2$, $\lambda_2=f_2g_2^2$, and $\lambda_3=f_3g_2^2$, and obtain the equations for them:
\begin{eqnarray}
\frac{df_1}{dG}&=&24f_1^2+\frac{9}{8}-(9+2b_2^\mathrm{s})f_1+\frac{5}{2}f_3^2,\\
\frac{df_2}{dG}&=&26f_2^2+108-(72+2b_2^\mathrm{s})f_2+2f_3^2,\\
\frac{df_3}{dG}&=&4f_3^2+18-\left(\frac{81}{2}+2b_2^\mathrm{s}\right)f_3+12f_1f_3+14f_2f_3.
\end{eqnarray}
where $G(t')=(b_2^\mathrm{s})^{-1}\ln[g_2(t')/g_2(0)]$.

In order to eliminate the linear terms, we further define functions $\hat{f}_1$ and $\hat{f}_2$ by the shifts $f_1=\hat{f}_1+1/8$ and $f_2=\hat{f}_2+69/52$. Then we have
\begin{equation}
\frac{d\hat{f}_1}{dG}=24\hat{f}_1^2+\frac{3}{4}+\frac{5}{2}f_3^2,\quad
\frac{d\hat{f}_2}{dG}=26\hat{f}_2^2+\frac{6471}{104}+2f_3^2.
\end{equation}
Noting that $\hat{f}_2$ runs much faster than $\hat{f}_1$ and $f_3$, we can simply neglect $f_3$ in the second equation and find its solution as
\begin{eqnarray}
\hat{f}_2(t')=\hat{d}\tan\left[\hat{c}_2\hat{d}G(t')+\tan^{-1}\left(\frac{\hat{f}_2(0)}{\hat{d}}\right)\right],
\end{eqnarray}
where $\hat{c}_2=26$ and $\hat{d}=\sqrt{6471/(104c_2)}\approx1.55$. The Landau pole is reached when $\hat{c}_2\hat{d}G(t')+\tan^{-1}[\hat{f}_2(0)/\hat{d}]=\pi/2$.
Thus, the corresponding scale is
\begin{equation}
\Lambda_\mathrm{LP}^{(f_2)} = \Lambda_\mathrm{s}\exp\left[-\frac{2\pi}{b_2^\mathrm{s}\alpha_2(\Lambda_\mathrm{s})}\left(\exp\left(-\frac{b_2^\mathrm{s}\pi}{\hat{c}_2\hat{d}}\left[1-\frac{2}{\pi}\tan^{-1}\left(\frac{\hat{f}_2(0)}{\hat{d}}\right)\right]\right)-1\right)\right].
\end{equation}

Setting $\hat{f}_2(0)=-69/52$, which corresponds to $\lambda_2=0$ at $\mu = \Lambda_\mathrm{s}$, we find that the maximal LP scale for $\lambda_2$ is $\Lambda_\mathrm{LP}^{(\lambda_2)}=5.6\times10^{14}~\si{GeV}$, which is far below the Planck scale.
Such a Landau pole implies that other new physics may exist between the quintuplet mass scale and the Planck scale, rendering all the couplings finite.

\section{5-5-1 model}\label{sec:551}

In this section, we will attempt to push up the LP scale obtained above.
A lesson we can learn from the standard model is that the top Yukawa coupling gives a negative contribution to the self-coupling of the Higgs boson. As the Landau pole is induced by the self-coupling of the quintuplet, it is straightforward to introduce extra fermions with a Yukawa coupling to the quintuplet to shift the Landau pole.
Such a motivation leads to the 5-5-1 model studied below.

\subsection{Yukawa interactions}

There are three minimal ways to construct Yukawa interactions with the quintuplet scalar: introducing fermions in $(\mathbf{1},0)\oplus(\mathbf{5},0)$, $(\mathbf{3},0)\oplus(\mathbf{3},0)$, and $(\mathbf{4},\pm1/2)\oplus(\mathbf{2},\mp1/2)$.
The first and second options have potential for explaining the tiny neutrino masses via the type-I and type-III seesaw mechanisms, respectively. In order to keep $\Delta^{(0)}$ stable, one of the two fermions participating the Yukawa interaction should be odd under the $Z_2$ symmetry. Particularly, in the second option, the two $(3,0)$ fermions should be different: one is $Z_2$-odd, and the other one is $Z_2$-even. Consequently, in order to give correct neutrino oscillation properties, we have to introduce at least one more triplet which is $Z_2$-even. On the other hand, in the third option, the $(\mathbf{4},\pm 1/2)$ representation should correspond to a $Z_2$-odd Dirac fermion for avoiding anomalies~\cite{Adler:1969gk,Bell:1969ts,Witten:1982fp,Bar:2002sa}. The $(\mathbf{2},\mp 1/2)$ fermions can just be the SM lepton doublets which are $Z_2$-even.
However, this case is less interesting for us, as it cannot explain the neutrino masses.

For these reasons, in this work we only concentrate on the $(\mathbf{1},0)\oplus(\mathbf{5},0)$ case. The resulting model is dubbed the ``5-5-1'' model. Minimally, we introduce a left-handed self-conjugated fermionic quintuplet $\Psi_\mathrm{L}$ and several right-handed fermionic singlets $N_{a,\mathrm{R}}$. It is convenient to use the tensor notation for writing down the interaction terms. The tensor notation can be translated to the familiar vector notation using the following dictionaries of $\Phi$ and $\Psi$:
\begin{equation}
\Phi=\frac{1}{\sqrt{2}}\begin{pmatrix}\Delta^{(2)}\\ \Delta^{(1)}\\ \Delta^{(0)}\\ \Delta^{(-1)}\\ \Delta^{(-2)}\\ \end{pmatrix}=\frac{1}{\sqrt{2}}\begin{pmatrix}\Phi_{1111}\\ 2\Phi_{1112}\\ \sqrt{6}\Phi_{1122}\\ -2\Phi_{1222}\\ \Phi_{2222}\end{pmatrix},\quad
\Psi_\mathrm{L}=\begin{pmatrix}\Psi_{+2,\mathrm{L}}\\ \Psi_{+1,\mathrm{L}}\\ \Psi_{0,\mathrm{L}}\\ \Psi_{-1,\mathrm{L}}\\ \Psi_{-2,\mathrm{L}}\end{pmatrix}=\begin{pmatrix}\Psi_{1111,\mathrm{L}}\\ 2\Psi_{1112,\mathrm{L}}\\ \sqrt{6}\Psi_{1122,\mathrm{L}}\\ -2\Phi_{1222,\mathrm{L}}\\ \Phi_{2222,\mathrm{L}}\end{pmatrix}.
\end{equation}
At the renormalizable level, the Yukawa interactions can be expressed as
\begin{equation}\label{yukawa}
\mathcal{L}_\mathrm{yuk}=
-(y_\nu)_{ab} \overline{\ell_{a,\mathrm{L}}^i}H_iN_{b,\mathrm{R}}
-\sqrt{6}y\Phi_{ijkl}\overline{\Psi_\mathrm{L}^{ijkl}}N_{3,\mathrm{R}}
+ \mathrm{h.c.},
\end{equation}
where $\ell_{a,\mathrm{L}}$ denotes the SM lepton doublets. $i,j,k,l=1,2$ are totally symmetric $\SUtwoL$ indices. $a$ and $b$ are family indices and at least two singlets are required for generating the realistic neutrino mixing.
Eq.~\eqref{yukawa} respects the $Z_2$ symmetry, with $\Psi_\mathrm{L}$ being $Z_2$-odd and $N_{a,\mathrm{R}}$ being $Z_2$-even.
Consequently, the new terms do not endanger stability of the scalar quintuplet, which has mass far below the fermionic one.

The above Yukawa interactions involve many new parameters. To illustrate the point, we adopt some working simplifications.
In the first, we assume that there are three generations of $N_R$, and the third generation has the largest Yukawa coupling, which is close to  the $\tau$ Yukawa coupling, \textit{i.e.}, $(y_\nu)_{33}\simeq y_\tau$.
Other elements in the $y_\nu$ matrix are much smaller but should be consistent with neutrino oscillation data.
Next, we assume all the new fermions, $\Psi_\mathrm{L}$ and $N_{a,\mathrm{R}}$ have the same mass $M_{51}$, which is a characteristic scale of the 5-5-1 model. Thus, the neutrino masses given by the seesaw mechanism are $\sim m_\tau^2/M_{51}$. According to the current cosmological constraint~\cite{Patrignani:2016xqp}, $m_\nu\sim m_\tau^2/M_{51}\lesssim0.2~\si{eV}$, which implies that $M_{51}\sim 10^{10}~\si{GeV}$. Hereafter, $M_{51}=10^{10}~\si{GeV}$ will be set as a benchmark scale of the 5-5-1 model.  In the concrete numerical analysis, we will comment on the situation of deviation from this scale setup. Finally, in the 5-5-1 Yukawa interaction term, \textit{i.e.}, the second term of $\mathcal{L}_\mathrm{yuk}$, we only consider the coupling to $N_{3,\mathrm{R}}$, neglecting the other two couplings. Thus, we just need to deal with one 5-5-1 Yukawa coupling, $y$. This 5-5-1 Yukawa term can be expanded as
\begin{eqnarray}
\mathcal{L}_\mathrm{yuk}&\supset&-\sqrt{3}y\big[\Delta^{(0)}\overline{\Psi_{0}}N_3+(\Delta^{(-1)}\overline{\Psi_{-1}}N_3+\Delta^{(-2)}\overline{\Psi_{-2}}N_3+\mathrm{h.c.})\big],
\end{eqnarray}
where $\Psi_{-Q}=\Psi_{-Q,\mathrm{L}}+(\Psi_{+Q,\mathrm{L}})^\mathrm{c}$ and $N_3=N_{3,\mathrm{R}}+(N_{3,\mathrm{R}})^\mathrm{c}$.
The gauge couplings of the fermionic quintuplet $\Psi$ are given by
\begin{eqnarray}
\mathcal{L}_{\Psi}&=&g_2\big(\sqrt{3}W^+_\mu\overline{\Psi_0}\gamma^\mu\Psi_{-1}+\sqrt{2}W^+_\mu\overline{\Psi_{-1}}\gamma^\mu\Psi_{-2}+\mathrm{h.c.}\big)\nonumber\\
&&-\big(eA_\mu + g_2 c_\mathrm{W} Z_\mu\big)
\big(2\overline{\Psi_{-2}}\gamma^\mu\Psi_{-2}+\overline{\Psi_{-1}}\gamma^\mu\Psi_{-1}\big).
\end{eqnarray}

\subsection{$\beta$ functions and analytic results}

The contributions of the quintuplet and singlet fermions at $\mu > M_{51}$ further modify the one-loop $\beta$ functions of the $\SUtwoL$ gauge coupling and scalar couplings:
\begin{eqnarray}
\delta \beta_{g_2}&=&\frac{1}{16\pi^2}\frac{20}{3}g_2^3,\quad
\delta \beta_{\lambda_2}=\frac{1}{16\pi^2}(-72y^4+16y^2\lambda_2),\\
\delta \beta_{\lambda_3}&=&\frac{1}{16\pi^2}8y^2\lambda_3,\quad
\beta_y =\frac{1}{16\pi^2}y(19y^2-18g_2^2)\label{betay}.
\end{eqnarray}
In these expressions, we have neglected the effect of the neutrino Yukawa couplings $(y_\nu)_{ab}$. This will be justified later.


Note that the $\beta$ function of $g_2$ becomes positive for $\mu>M_{51}$. Thus, one may worry about up to which scale $g_2$ can remain perturbative.
By requiring $\alpha_2=4\pi$, we find the non-perturbative scale of $g_2$ in the 5-5-1 model as
\begin{eqnarray}
\Lambda_{g_2}^\mathrm{NP}=M_{51}\left(\frac{\Lambda_\mathrm{s}}{M_{51}}\right)^{b_2^\mathrm{s}/b_2^\mathrm{tot}}\exp\left[\frac{2\pi}{b_2^\mathrm{tot}}\left(\alpha_2^{-1}(\Lambda_\mathrm{s})-\frac{1}{4\pi}\right)\right],
\end{eqnarray}
where $b_2^\mathrm{tot}=b_2^\mathrm{s}+20/3=31/6$.
Setting $\Lambda_\mathrm{s}=10~\si{TeV}$, we find that almost any $M_{51}>\Lambda_\mathrm{s}$ will give $\Lambda_{g_2}^\mathrm{NP}>M_\mathrm{Pl}$.
Thus, $g_2$ would still be perturbative at the Planck scale.

By solving the RGEs, we obtain the exact values of $g_2(\mu)$ and $y(\mu)$ at $\mu>M_{51}$ as
\begin{eqnarray}
g_2^2(\mu)&=&g_2^2(M_{51}) \left[1-\frac{1}{8\pi^2}b_2^\mathrm{tot}g_2^2(M_{51})\ln\frac{\mu}{M_{51}}\right]^{-1},\\
y^2(\mu)&=&(18+b_2^\mathrm{tot})g_2^2(\mu)
\left[19+F_0\left(\frac{g_2^2(\mu)}{g_2^2(M_{51})}\right)^{(18+b_2^\mathrm{tot})/b_2^\mathrm{tot}}\right]^{-1},
\end{eqnarray}
where
\begin{equation}
F_0 \equiv (18+b_2^\mathrm{tot})\frac{g_2^2(M_{51})}{y^2(M_{51})}-19.\label{F0def}
\end{equation}
$F_0 = 0$ leads to a critical value of $y(M_{51})$,
\begin{equation}
y_\mathrm{c}(M_{51}) \equiv
\sqrt{\frac{18+b_2^\mathrm{tot}}{19}} g_2(M_{51}).
\end{equation}

If $y(M_{51})>y_c(M_{51})$, \textit{i.e.}, $F_0<0$, the 5-5-1 Yukawa coupling $y$ will reach a Landau pole at
\begin{eqnarray}
\Lambda_\mathrm{LP}^{(y)}=M_{51}\exp\left[\frac{8\pi^2}{b_2^\mathrm{tot}g_2^2(M_{51})}\left(1-\left[1-\frac{18+b_2^\mathrm{tot}}{19}\frac{g_2^2(M_{51})}{y^2(M_{51})}\right]^{b_2^\mathrm{tot}/(18+b_2^\mathrm{tot})}\right)\right].
\end{eqnarray}
Then the condition for $\Lambda_\mathrm{LP}^{(y)}>M_\mathrm{Pl}$ is
\begin{eqnarray}
y^2(M_{51})<\frac{18+b_2^\mathrm{tot}}{19}g_2^2(M_{51})
\left[1-\left(1-\frac{b_2^\mathrm{tot}g_2^2(M_{51})}{8\pi^2}\ln\frac{M_\mathrm{Pl}}{M_{51}}\right)^{(18+b_2^\mathrm{tot})/b_2^\mathrm{tot}}\right]^{-1}.
\end{eqnarray}
Note that the perturbative condition $y\leq4\pi$ will give a smaller upper bound on $y(M_{51})$.

As $y$ grows at high scales, the effect of $(y_\nu)_{ab}$ may become important.
The $\beta$ function of $y_\nu$ with all Yukawa coupling included is
\begin{equation}
\beta_{y_\nu}=\frac{y_\nu}{16\pi^2}\left[\frac{3}{2}y_\nu^\dag y_\nu+T+\frac{15}{2}y^2-\frac{3}{2}y_e^\dag y_e-\frac{9}{4}g_2^2-\frac{9}{20}g_1^2\right],
\end{equation}
where $T=3\mathrm{tr}(y_u^\dag y_u)+3\mathrm{tr}(y_d^\dag y_d)+\mathrm{tr}(y_e^\dag y_e)+\mathrm{tr}(y_\nu^\dag y_\nu)$, and $y_u$, $y_d$, $y_e$, $y_\nu$ are the Yukawa coupling matrices for quarks and leptons.
For simplicity we consider that only one real element of $y_\nu$ on the diagonal dominates, and denote it $\hat{y}_\nu$.
At high scales, $y$ and $g_2$ are large, and the above equation can be approximated as
\begin{eqnarray}
\beta_{\hat{y}_\nu}\simeq\frac{\hat{y}_\nu}{16\pi^2}\left[\frac{5}{2}\hat{y}_\nu^2+\frac{15}{2}y^2-\frac{9}{4}g_2^2\right].
\end{eqnarray}
As $\hat{y}_\nu$ contributes to the self-energy of $N_{3,\mathrm{R}}$, it will modify $\beta_y$ by
\begin{eqnarray}
\delta\beta_{y}=\frac{y}{16\pi^2}\hat{y}_\nu^2.
\end{eqnarray}
Therefore, the growing $\hat{y}_\nu$ will boost the running of $y$, and hence the LP scale of $y$ becomes lower.
Nonetheless, $\hat{y}_\nu$ is important only when $y$ is very large (not far from its LP), so the LP scale would not change too much.
Thus, it is still reasonable to neglect the effect of $(y_\nu)_{ab}$.

On the other hand, if $y(M_{51})< y_\mathrm{c}(M_{51})$, \textit{i.e.}, $F_0>0$, as $\mu$ goes up $y(\mu)$ will increase at the beginning, and then turn its direction at some scale, and then exponentially drop down to zero.
If the decreasing behavior happens at a scale lower than the Planck scale, the effect of the 5-5-1 Yukawa coupling is not significant. In this case, the model is quite similar to the original quintuplet MDM model, where $\lambda_2$ blows up before the Planck scale.
If the decrease happens at some scale higher than the Planck scale, then $y$ might be large enough to keep $\lambda_2$ finite.

\subsection{Numerical calculation}\label{subsect3.3}

The above analysis is based on analytic calculations. Below we present the results obtained by solving the RGEs numerically.

\begin{figure}[!t]
\centering
\subfigure[~$y(M_{51})=0.662$.]
{\includegraphics[width=0.48\textwidth]{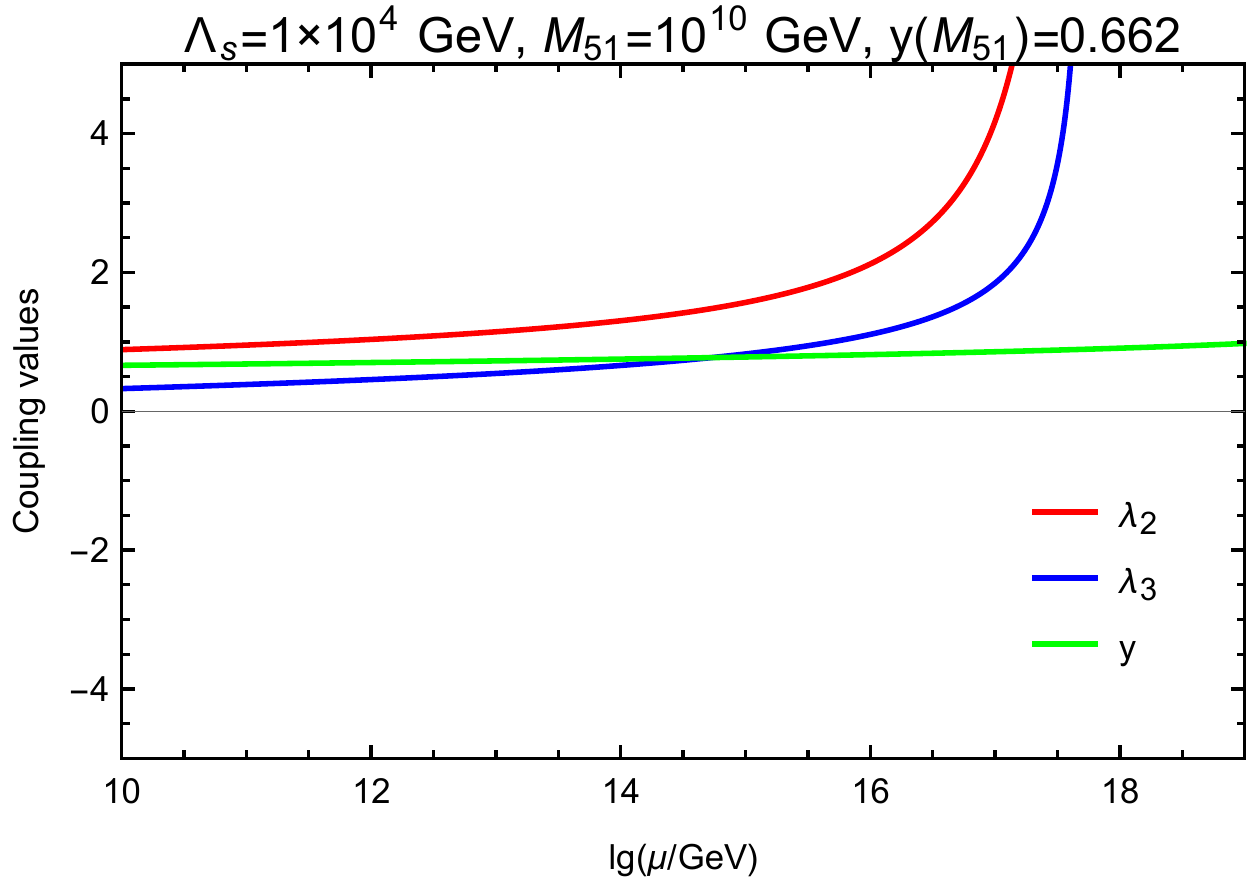}}
\hspace{.01\textwidth}
\subfigure[~$y(M_{51})=0.666$.]
{\includegraphics[width=0.48\textwidth]{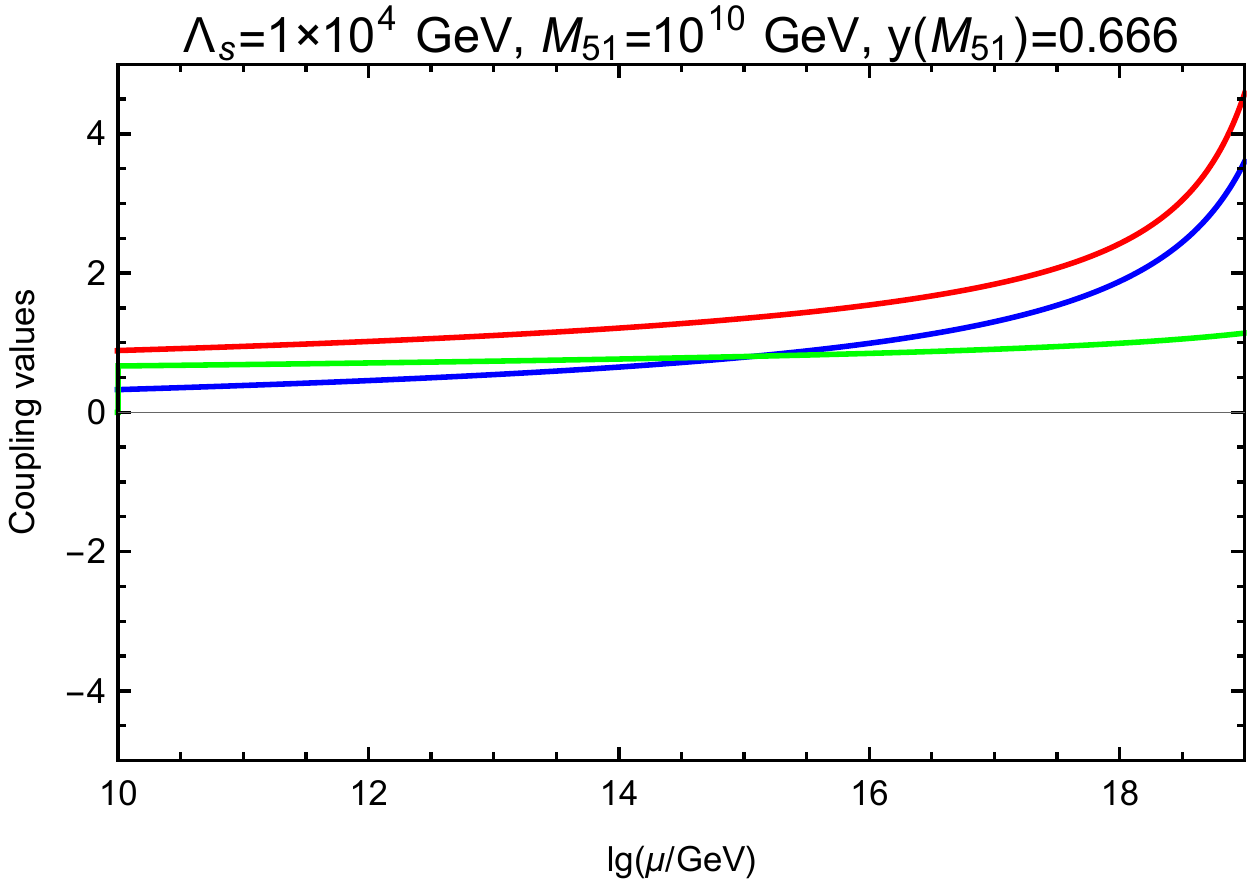}}
\subfigure[~$y(M_{51})=0.667$.]
{\includegraphics[width=0.48\textwidth]{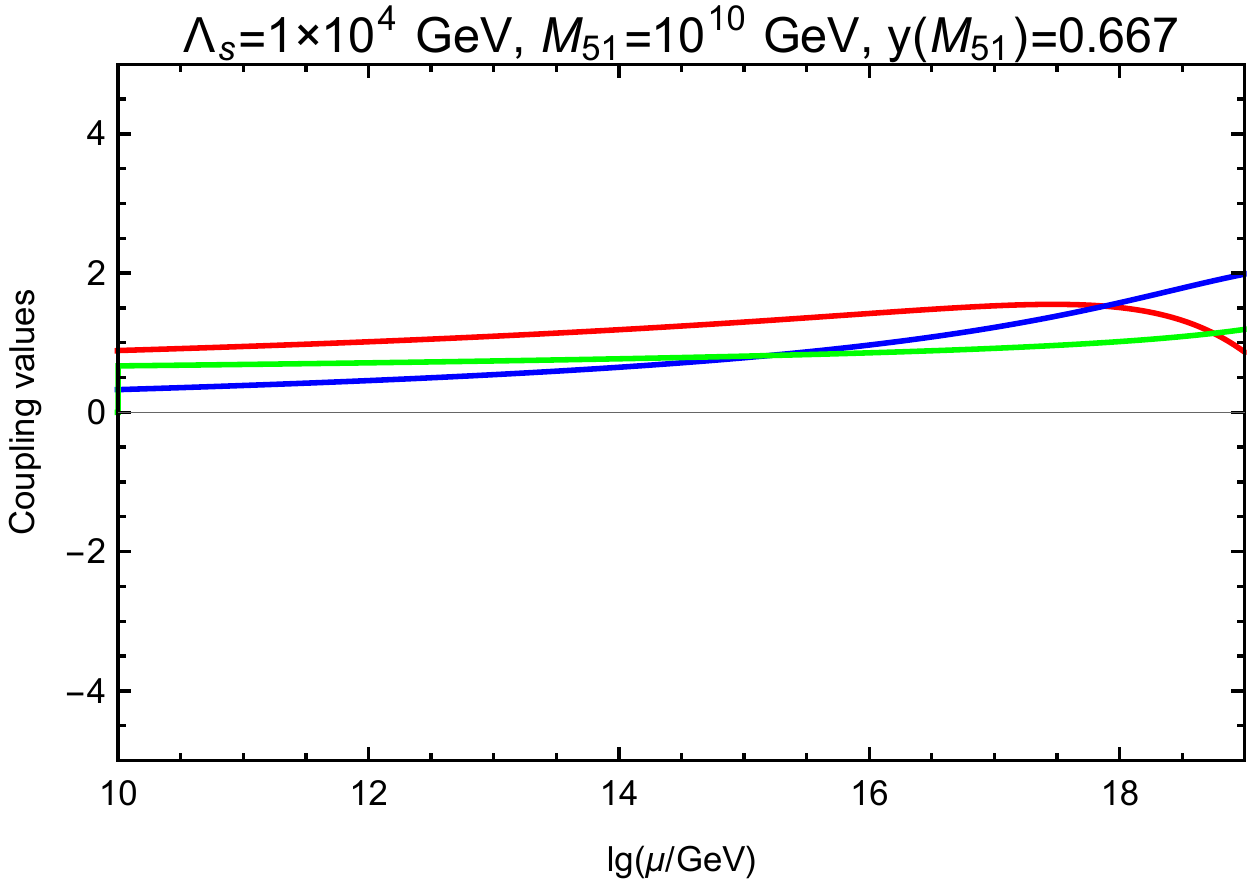}}
\hspace{.01\textwidth}
\subfigure[~$y(M_{51})=0.672$.]
{\includegraphics[width=0.48\textwidth]{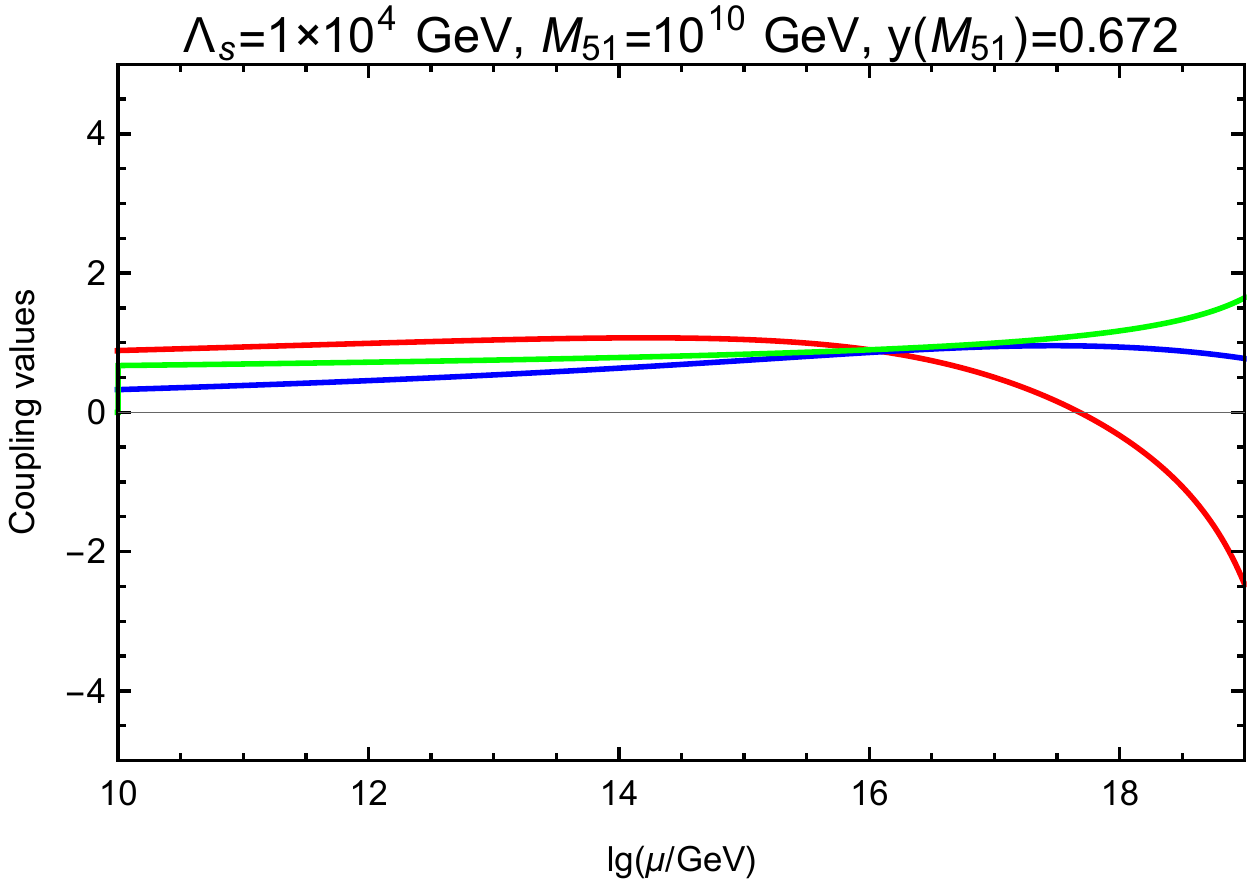}}
\caption{Evolution of the couplings $\lambda_2$, $\lambda_3$, and $y$ in the 5-5-1 model with different values of $y(M_{51})$. We have fixed $\Lambda_\mathrm{s}=10~\si{TeV}$, $M_{51}=10^{10}~\si{GeV}$, $\lambda_2(\Lambda_\mathrm{s})=0.015$, and $\lambda_3(\Lambda_\mathrm{s})=0.2$. The red, blue, and green lines correspond to the evolution of $\lambda_2$, $\lambda_3$, and $y$, respectively.}\label{fig:evolution}
\end{figure}

Firstly, we investigate the impact of $y(M_{51})$ on the running of couplings.
Fig.~\ref{fig:evolution} shows the evolution of the couplings with different values of $y(M_{51})$ for fixing $\Lambda_\mathrm{s}=10$~TeV,  $M_{51}=10^{10}$~GeV, $\lambda_2(\Lambda_\mathrm{s})=0.015$, and $\lambda_3(\Lambda_\mathrm{s})=0.2$.
We can see that the evolution behavior dramatically depends on the delicate input value of $y(M_{51})$.
\begin{itemize}
\item If $y(M_{51})=0.662$, it will be unable to slow down the growing of $\lambda_2$, which reaches a Landau pole at a scale lower than the Planck scale.
\item If $y(M_{51})=0.666$, the Landau pole scale of $\lambda_2$ will be push up to near the Planck scale.
\item If $y(M_{51})=0.667$, all couplings will remain perturbative up to the Planck scale, and the VS conditions will be satisfied at the same time.
\item If $y(M_{51})=0.672$, although all couplings will remain perturbative up to the Planck scale, $\lambda_2$ will become negative, leading to an unstable vacuum before the Planck scale.
\end{itemize}
The above fine-tuning of $y(M_{51})$ is expected. In the $\beta$ function of $\lambda_2$ one has to arrange a delicate cancellation between $108g_2^4$ and $72y^4$ at $M_{51}$ so that $\beta_{\lambda_2}$ is well under control in a wide energy region. Otherwise, the effect of Yukawa damping on $\lambda_2$ is insufficient for a slightly smaller $y(M_{51})$ and too much for a slightly larger $y(M_{51})$ that renders a negative $\beta_{\lambda_2}$ too early.

\begin{figure}[!t]
\centering
\subfigure[~$y(M_{51})=0.662$.]
{\includegraphics[width=0.42\textwidth]{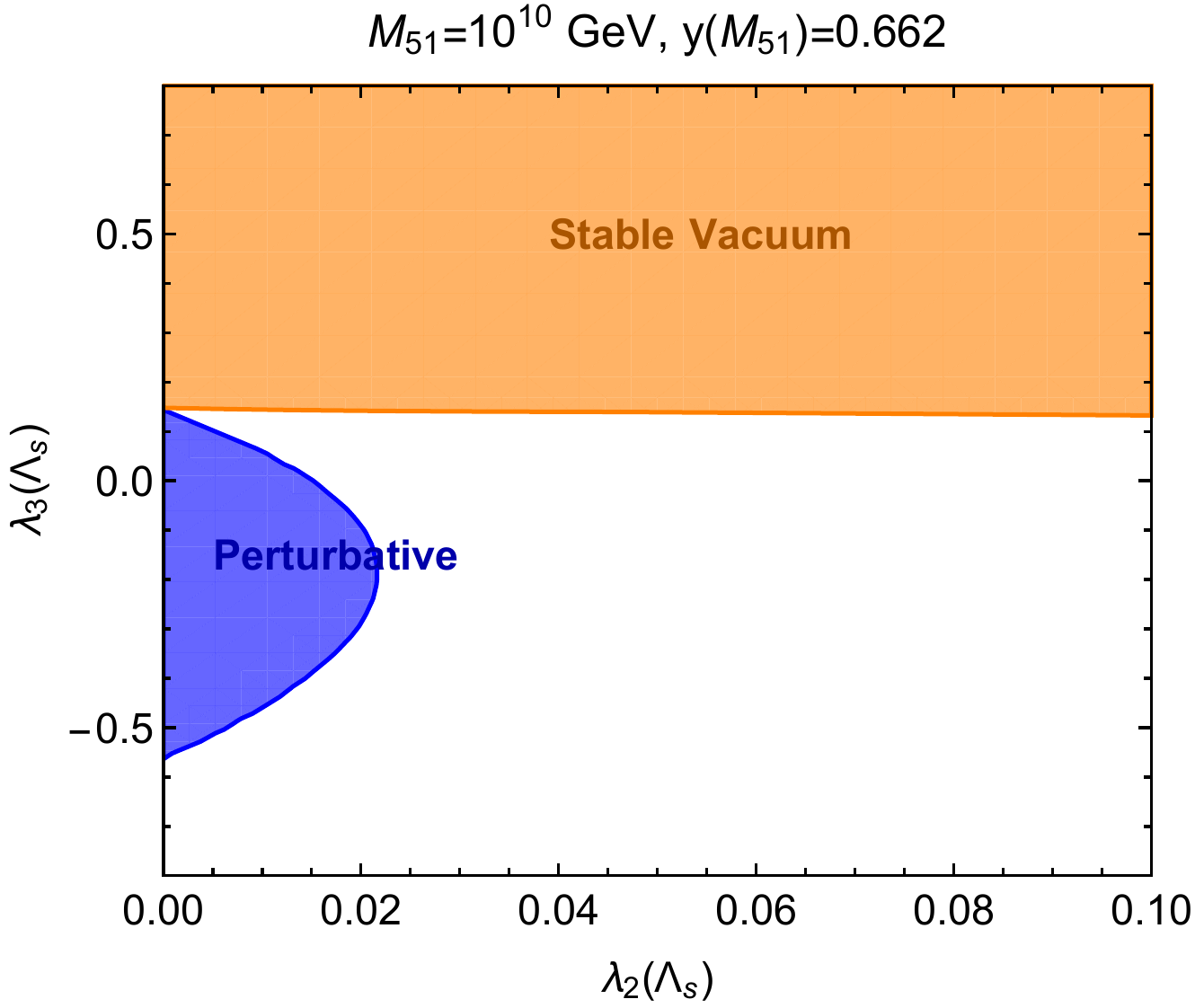}} \subfigure[~$y(M_{51})=0.666$.]
{\includegraphics[width=0.42\textwidth]{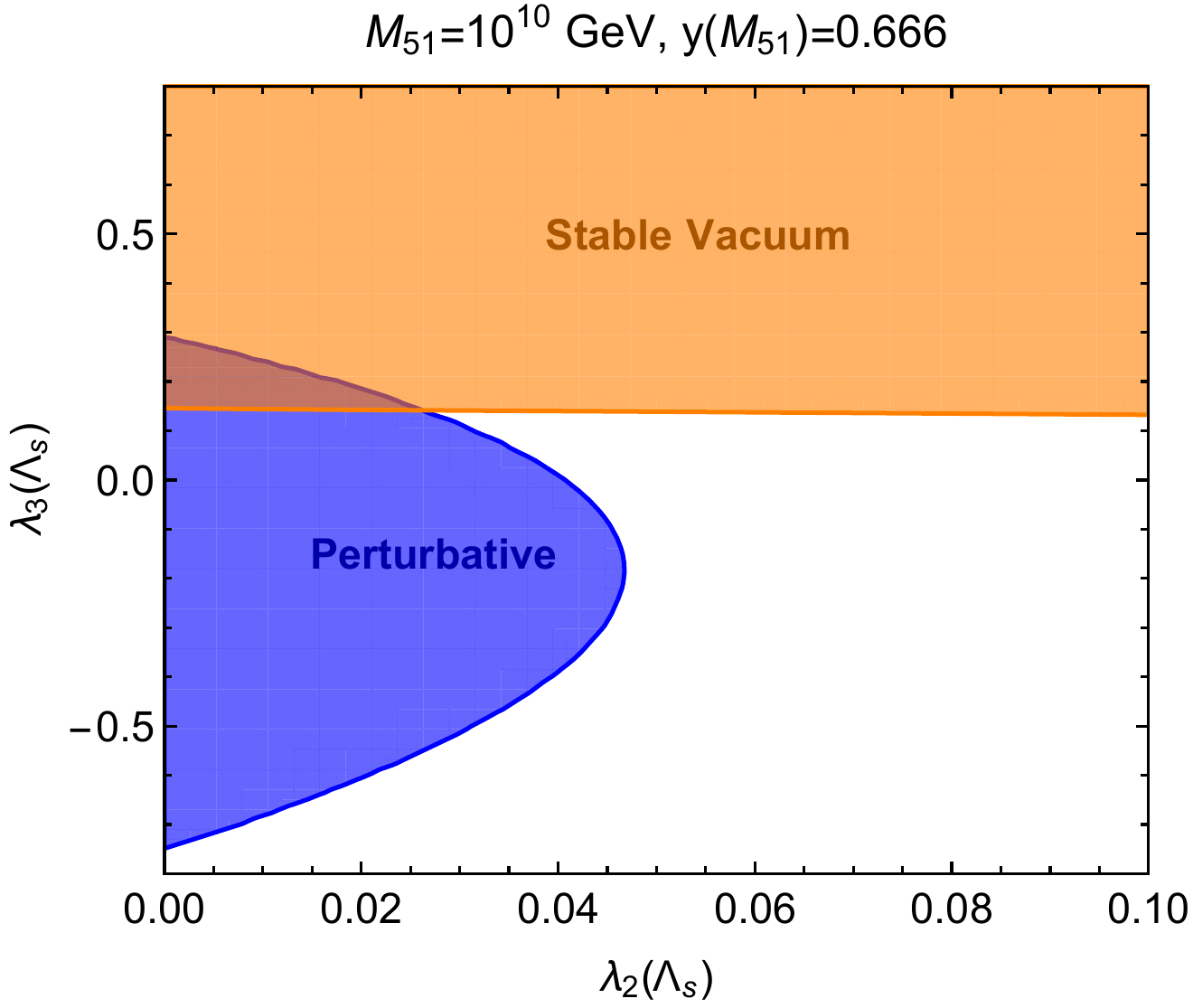}}
\subfigure[~$y(M_{51})=0.670$.]
{\includegraphics[width=0.42\textwidth]{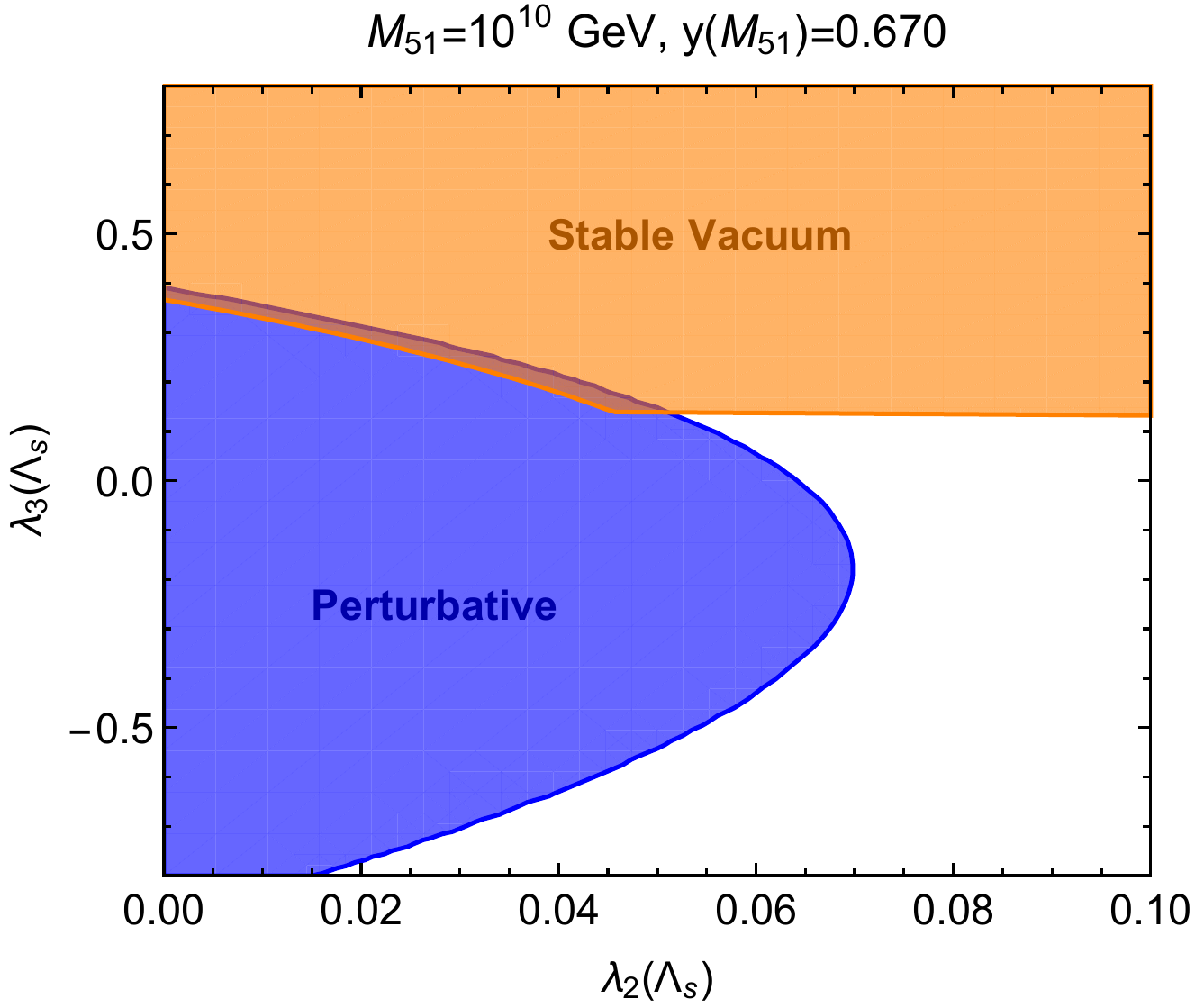}}
\caption{Regions favored by the perturbativity and VS conditions in
the $\lambda_2(\Lambda_\mathrm{s})$-$\lambda_3(\Lambda_\mathrm{s})$ plane for the 5-5-1 model with different values of $y(M_{51})$. We have fixed  $\Lambda_\mathrm{s}=10~\si{TeV}$ and $M_{51}=10^{10}~\si{GeV}$. The blue and orange regions correspond to the parameter regions satisfying the perturbativity and VS conditions, respectively. The overlap regions are favored by both conditions.}\label{fig:Pert_VS}
\end{figure}

Secondly, we study the parameter regions where the perturbativity and VS conditions are satisfied.
We choose $y(M_{51})=0.662,0.666,0.670$ as three typical inputs, and perform scans in the $\lambda_2$-$\lambda_3$ plane with $\Lambda_\mathrm{s}=10~\si{TeV}$ and $M_{51}=10^{10}~\si{GeV}$.
The results are shown in Fig.~\ref{fig:Pert_VS}.
In the blue regions, all parameters can remain perturbative up to the Planck scale, while in the orange region, the vacuum remains stable up to the Planck scale.
The overlap regions simultaneously satisfy the perturbativity and VS conditions.

As we vary the initial value of the Yukawa coupling $y(M_{51})$, the overlap region varies. We find that the perturbativity and VS conditions constrain $\lambda_3(\Lambda_\mathrm{s})$ within a range of $0.14<\lambda_3(\Lambda_\mathrm{s})<0.4$, and $\lambda_2(\Lambda_\mathrm{s})$ within a range of $0<\lambda_2(\Lambda_\mathrm{s})<0.053$.
This favored range of $\lambda_3(\Lambda_\mathrm{s})$ is also indicated as a green strip in Fig.~\ref{fig:constr} for comparing with other phenomenological constraints.

\begin{figure}[!t]
\centering
\includegraphics[width=.5\textwidth]{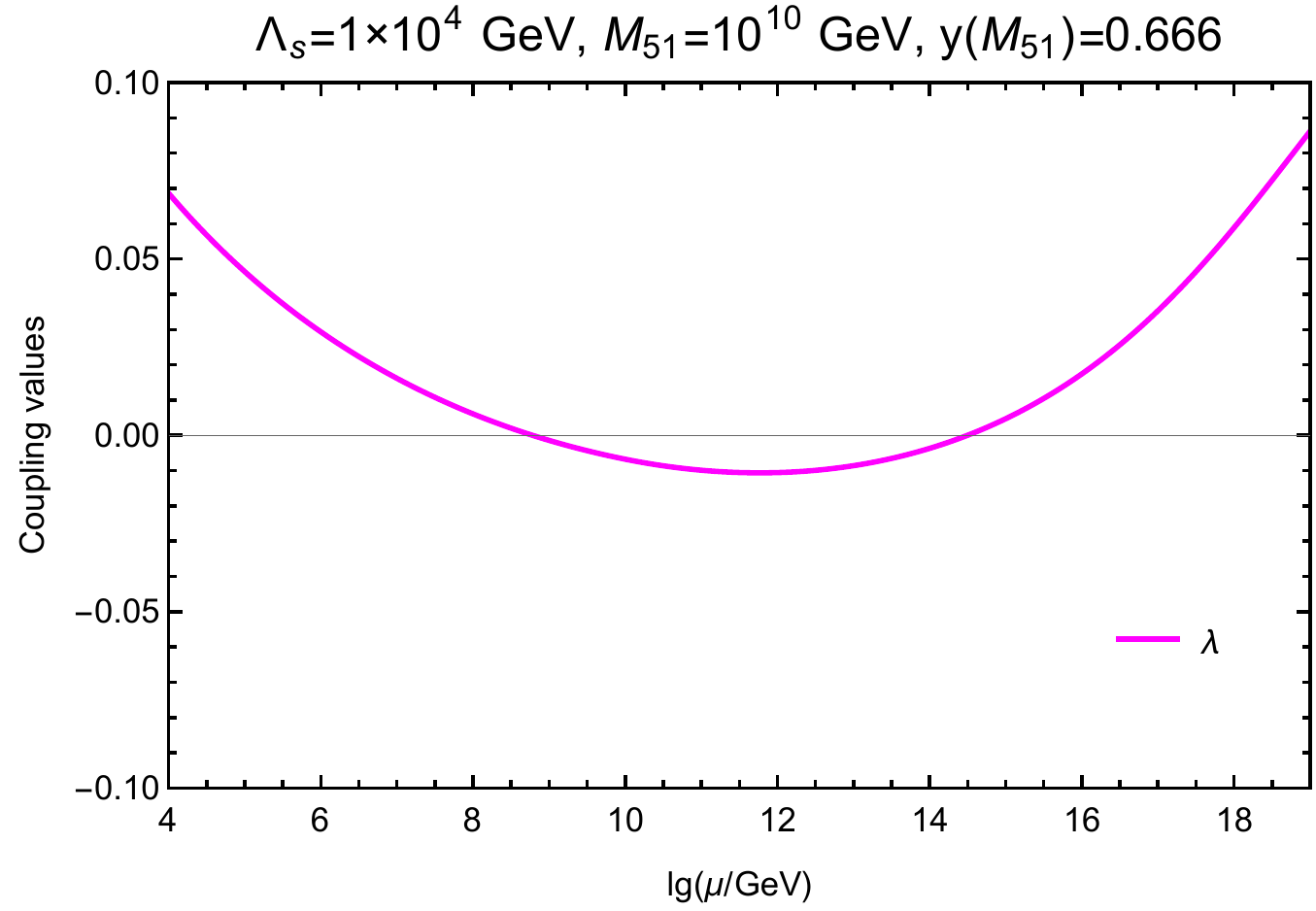}
\caption{Evolution of $\lambda$ in the 5-5-1 model with $\Lambda_\mathrm{s}=10~\si{TeV}$, $M_{51}=10^{10}~\si{GeV}$,  $\lambda_2(\Lambda_\mathrm{s})=0.015$, $\lambda_3(\Lambda_\mathrm{s})=-0.02$, and $y(M_{51})=0.666$.}\label{fig:lambda}
\end{figure}

The VS condition $\lambda_3+2\sqrt{\lambda\lambda_2}\geq0$ seems to allow a small negative $\lambda_3(\Lambda_\mathrm{s})$.
However, the results in Fig.~\ref{fig:Pert_VS} exclude the whole regions with negative $\lambda_3(\Lambda_\mathrm{s})$.
The reason is that the vacuum stability also requires the Higgs quartic coupling $\lambda$ remaining positive when running to higher scales.
In the SM, $\lambda$ will turn negative at a scale $\sim 10^{9}~\si{GeV}$.
The existence of the quintuplet scalar could change this behavior, because the $\lambda_3$ coupling has a positive contribution to the $\beta$ function of $\lambda$, as shown in Eq.~\eqref{betafunc_2}.
If $|\lambda_3|$ is not large enough to turn around the trend of $\lambda$, however, the vacuum will be unstable at some scale.
We set $\lambda_3(\Lambda_\mathrm{s})=-0.02$, $\lambda_2(\Lambda_\mathrm{s})=0.015$, and $y(M_{51})=0.666$, and demonstrate the evolution of $\lambda(\mu)$ in Fig.~\ref{fig:lambda}.
We can see that $\lambda$ goes to negative values at scales  $\sim 10^9-10^{14}~\si{GeV}$.
In this case, the vacuum stability cannot be ensured along the whole way to the Planck scale.

To end up this section, we would like to make a comment on the possible impacts from a different scale set for $M_{51}$ other than the benchmark scale $10^{10}$ GeV. If $M_{51}$ lies below this scale, the Yukawa term can slow down the running of $\lambda_2$ because of the lower scale, thus requiring a smaller $y(M_{51})$. But $M_{51}$ cannot be many orders of magnitude below, because $\beta_{\lambda_2}$ is about to change the sign and render $\lambda_2$ negative even far below the Planck scale. $M_{51}\gg 10^{10}$ GeV is also disfavored, because $\lambda_2(M_{51})$ would be too large to be stopped by a perturbative $y$. One can employ a parallel  analysis for other choice of  $M_{51}$, and the configuration of the viable parameter space will not change significantly as long as $M_{51}$ is not very far from the benchmark value.

\section{Discussion and conclusions}\label{sect4}

Perturbativity puts a strong constraint on scalar MDM models, especially when the multiplet lives in a large $\SUtwoL$ representation. The scalar self-couplings usually reach a Landau pole at a energy scale far below the Planck scale, in spite of the initial values.
There are two reasons leading to such a disaster.
One is that the quadratic self-interaction Lagrangians result in terms with large coefficients in the $\beta$ functions of the self-couplings.
Once the self-couplings obtains a modest value, such terms will drive the self-couplings to grow exponentially and soon violate
perturbativity.
Another reason is that there are also significant terms contributed by other couplings in the $\beta$ functions of the self-couplings, \textit{e.g.}, a $g_2^4$ term with a large coefficient.
These terms ensure that even when their initial values are very tiny, the self-couplings will quickly obtain modest values after a short journey of running.

In a previous work~\cite{Cai:2015kpa}, we found that the perturbativity problem in the real septuplet scalar MDM model is quite stubborn: even after some fermions with Yukawa interactions are introduced to slow down the running of scalar self-couplings, the model is still unable to remain perturbative up to the Planck scale.
Nonetheless, this may be achievable if the MDM scalar lives in a smaller representation.
Therefore, we have studied the real scalar quintuplet MDM model in this work.
The observed relic abundance implies that the scale for introducing such a quintuplet should be near $10~\si{TeV}$. Our calculation suggests that the quintuplet self-coupling $\lambda_2$ will hit a Landau pole at a scale of $5.6\times10^{14}~\si{GeV}$, which is consistent with Ref.~\cite{Hamada:2015bra}.

In order to push up this LP scale, we have extended the model with Yukawa couplings of the scalar quintuplet to a fermionic quintuplet and three fermionic singlets, resulting in the so-called 5-5-1 model.
The new singlets can also play the role of right-handed neutrinos, explaining the smallness of neutrino masses by the type-I seesaw mechanism.
We have found that if such Yukawa couplings are involved after a scale of $M_{51}\sim 10^{10}~\si{GeV}$, all couplings can remain perturbative up to the Planck scale,
The reason is that the Yukawa couplings contribute a large negative term to the $\beta$ function of $\lambda_2$ and hence slow down the growing of $\lambda_2$ at high scales.

We have also investigated the parameter regions favored by the perturbativity and vacuum stability conditions up to the Planck scale.
It has been found that these conditions constrain the Higgs-quintuplet coupling $\lambda_3(\Lambda_\mathrm{s})$ at the quintuplet scale $\Lambda_\mathrm{s}$ within a range of $0.14<\lambda_3(\Lambda_\mathrm{s})<0.4$, and the quintuplet self-coupling $\lambda_2(\Lambda_\mathrm{s})$ within a range of $0<\lambda_2(\Lambda_\mathrm{s})<0.53$.

\begin{acknowledgments}
This work is supported by the National Natural Science Foundation of China (NSFC)
under Grant Nos. 11375277, 11410301005, 11647606, 11005163, 11775086, 11875327, and 11805288, the Fundamental Research
Funds for the Central Universities,
the Natural Science Foundation of Guangdong Province under Grant No. 2016A030313313,
and the Sun Yat-Sen University Science Foundation.
\end{acknowledgments}

\appendix

\section{$\beta$ functions in the SM and initial values of couplings}\label{app:beta}

At one-loop level, the $\beta$ functions of SM couplings are give by
\begin{eqnarray}
&&\beta_{g_1}^\mathrm{SM}=\frac{1}{16\pi^2}\frac{41}{10}g_1^3,~~
\beta_{g_2}^\mathrm{SM}=\frac{1}{16\pi^2}\left(-\frac{19}{6}\right)g_2^3,~~
\beta_{g_3}^\mathrm{SM}=\frac{1}{16\pi^2}(-7)g_3^3,
\label{SM:betafunc_1}\\
&&\beta_{y_t}^\mathrm{SM}=\frac{1}{16\pi^2}y_t\left(\frac{9}{2}y_t^2-\frac{9}{4}g_2^2-\frac{17}{20}g_1^2-8g_3^2\right),
\label{SM:betafunc_2}\\
&&\beta_{\lambda}^\mathrm{SM}=\frac{1}{16\pi^2}\left\{24\lambda^2-6y_t^4+\frac{3}{8}\left[2g_2^4+\left(g_2^2+\frac{3}{5}g_1^2\right)^2\right]+\lambda\left(-9g_2^2-\frac{9}{5}g_1^2+12y_t^2\right)\right\}.
\label{SM:betafunc_3}
\end{eqnarray}
As most of the Yukawa couplings are negligible, only the top Yukawa coupling $y_t$ is considered in the above expressions.

In the RGE calculation, we use the following $\overline{\mathrm{MS}}$ values at $m_Z$ as initial values for gauge couplings:~\cite{Beringer:1900zz}
\begin{eqnarray}
\alpha_\mathrm{s}(m_Z)&=&\frac{1}{4\pi}[g_3(m_Z)]^2=0.1184,\\
\alpha(m_Z)&=&\frac{1}{4\pi}[g_2(m_Z)s_\mathrm{W}(m_Z)]^2=\frac{1}{127.926},\\
s_\mathrm{W}^2&=&\sin^2\theta_\mathrm{W}(m_Z)=0.2312.
\end{eqnarray}
The measured values of $y_t$ and $\lambda$ are obtained from the pole masses of the top quark and the Higgs boson, $m_t$ and $m_h$, respectively.
Therefore, we need to derive their $\overline{\mathrm{MS}}$ values $y_t(m_Z)$ and $\lambda(m_Z)$ at $m_Z$ by the matching conditions~\cite{Hambye:1996wb}
\begin{equation}
y_t(\mu_0)=\frac{\sqrt{2}m_t}{v}[1+\delta_t(\mu_0)],~~
\lambda(\mu_0)=\frac{m_h^2}{2v^2}[1+\delta_h(\mu_0)],
\end{equation}
setting $\mu_0 = m_Z$.
The related functions are
\begin{eqnarray}
\delta_t(\mu_0)&=&\left(-\frac{4\alpha_s}{4\pi}-\frac{4}{3}\frac{\alpha}{4\pi}+\frac{9}{4}\frac{m_t^2}{16\pi^2v^2}\right)\ln\frac{\mu_0^2}{m_t^2}+c_t,\\
\delta_h(\mu_0)&=&\frac{2v^2}{m_h^2}\frac{1}{32\pi^2v^4}[h_0(\mu_0)+m_h^2h_1(\mu_0)+m_h^4h_2(\mu_0)],\\
h_0(\mu_0)&=&-24m_t^4\ln\frac{\mu_0^2}{m_t^2}+6m_Z^4\ln\frac{\mu_0^2}{m_Z^2}+12m_W^4\ln\frac{\mu_0^2}{m_W^2}+c_0,\\
h_1(\mu_0)&=&12m_t^2\ln\frac{\mu_0^2}{m_t^2}-6m_Z^2\ln\frac{\mu_0^2}{m_Z^2}-12m_W^2\ln\frac{\mu_0^2}{m_W^2}+c_1,\\
h_2(\mu_0)&=&\frac{9}{2}\ln\frac{\mu_0^2}{m_h^2}+\frac{1}{2}\ln\frac{\mu_0^2}{m_Z^2}+\ln\frac{\mu_0^2}{m_W^2}+c_2.
\end{eqnarray}
Here the constants $c_0$, $c_1$, and $c_2$ are independent of $\mu_0$. Their contributions to $\delta_h$ are less than 0.02 and can be neglected. The constant $c_t$ lies in a range of $-0.052\leq c_t\leq-0.042$. We take $c_t=-0.052$ in the calculation, but choosing another value within the range would not essentially change our results.

\section{Sommerfeld enhancement effect}
\label{app:SE}

In order to account for the SE effect on the DM relic abundance and the annihilation cross section related in indirect detection, we consider the two-body Schr\"{o}dinger equations for (co-)annihilation pairs of dark sector particles, following  Refs.~\cite{Cirelli:2007xd,Cirelli:2015bda}. The two-body states can be categorized by their charges, denoted as $Q$. For the scalar quintuplet, there are three types of initial states ($Q=0,1,2$), and each of them is endowed with an annihilation rate matrix and a potential matrix.
The potential matrices account for the non-relativistic effect of exchanging electroweak gauge bosons between the two incoming particles, with the following forms:
\begin{eqnarray}\label{SEpotential}
V_{Q=0}&=&\begin{pmatrix}8\Delta m-4A&-2B&0\\-2B&2\Delta m-A&-3\sqrt{2}B\\0&-3\sqrt{2}B&0\end{pmatrix},\nonumber\\
V_{Q=1}&=&\begin{pmatrix}5\Delta m-2A&-\sqrt{6}B\\-\sqrt{6}B&2\Delta m-3B\end{pmatrix},\\
V_{Q=2}&=&\begin{pmatrix}4\Delta m&-2\sqrt{3}B\\-2\sqrt{3}B&2\Delta m+A\end{pmatrix}.\nonumber
\end{eqnarray}
Here $A=\alpha/r+\alpha_2c_\mathrm{W}^2e^{-m_Zr}/r$ and  $B=\alpha_2e^{-m_Wr}/r$ corresponds to neutral and charged gauge bosons, respectively.

The Schr\"{o}dinger equations for the two-body wave functions $\psi_i^{(j)}$ are given by
\begin{eqnarray}
-\frac{1}{m_0}\frac{\partial^2\psi_i^{(j)}}{\partial r^2}+\sum\limits_{k} V_{ik}\psi_{k}^{(j)}=K\psi_i^{(j)},
\end{eqnarray}
where $K=m_0\beta^2$ with $\beta$ a characteristic velocity of DM particles.
We adopt $\beta=10^{-3}$ and $0.28$ for DM particles in the Galaxy and at the freeze-out epoch, respectively. 
The boundary conditions are
\begin{eqnarray}
\psi_{i}^{(j)}(0)=\delta_{i}^j,\quad \frac{\partial\psi_i^{(j)}}{\partial r}(\infty)=i\sqrt{m_0(K-V_{ii}(\infty))}\,\psi_i^{(j)}(\infty).
\end{eqnarray} 
After solving these equations, annihilation and coannihilation cross sections with the SE effect are given by
\begin{eqnarray}
\sigma_i v=(A\Gamma A^\dag)_{ii}
\end{eqnarray} 
where $A_{ij}=\psi_i^{(j)}(\infty)$, and $\Gamma$ is the tree-level annihilation rate matrix.

The annihilation rate matrices account for annihilation and coannihilation of dark sector particles into SM particles at tree level.
Dominant final states are pairs of gauge bosons and of Higgs bosons. Using the optical theorem, the annihilation rate matrices are given by
\begin{eqnarray}
\Gamma_{Q=0}&=&\frac{6\pi\alpha_2^2}{25m_0^2}\begin{pmatrix}12&6&2\sqrt{2}\\6&9&5\sqrt{2}\\ 2\sqrt{2}&5\sqrt{2}&6\end{pmatrix}+\frac{\lambda_3^2}{200\pi m_0^2}\begin{pmatrix}1&1&\frac{1}{\sqrt{2}}\\1&1&\frac{1}{\sqrt{2}}\\ \frac{1}{\sqrt{2}}&\frac{1}{\sqrt{2}}&\frac{1}{2}\end{pmatrix}, \nonumber\\
\Gamma_{Q=1}&=&\frac{6\pi\alpha_2^2}{25m_0^2}\begin{pmatrix}6&\sqrt{6}\\ \sqrt{6}&1\end{pmatrix}, \\
\Gamma_{Q=2}&=&\frac{6\pi\alpha_2^2}{25m_0^2}\begin{pmatrix}4&-\sqrt{12}\\ -\sqrt{12}&3\end{pmatrix}.\nonumber
\end{eqnarray}
These matrices are utilized in the calculation of relic abundance.
One may check that the sum of all the diagonal elements can reproduce the tree-level effective annihilation cross section \eqref{eq:sv_eff}. Note that when computing annihilation cross sections, the states with  $Q=1,2$ contribute twice since there are contributions from the $(++,-)$ pair as well as from the $(--,+)$ pair.

In order to estimate the constraint from the MAGIC and Fermi-LAT experiments whose result was obtained by assuming DM totally annihilating into $W^+W^-$~\cite{Ahnen:2016qkx},
we also utilize the annihilation rate matrix for $\Delta^{(0)}\Delta^{(0)}\rightarrow W^+W^-$,
\begin{eqnarray}
\Gamma^{Q=0}_{WW}&=&\frac{2\pi\alpha_2^2}{m_0^2}\begin{pmatrix}4&10&6\sqrt{2}\\10&25&15\sqrt{2}\\ 6\sqrt{2}&15\sqrt{2}&18\end{pmatrix}+\frac{\lambda_3^2}{16\pi m_0^2}\begin{pmatrix}1&1&\frac{1}{\sqrt{2}}\\1&1&\frac{1}{\sqrt{2}}\\ \frac{1}{\sqrt{2}}&\frac{1}{\sqrt{2}}&\frac{1}{2}\end{pmatrix}.
\end{eqnarray}

\bibliographystyle{utphys}
\bibliography{ref}

\providecommand{\href}[2]{#2}\begingroup\raggedright\begin{thebibliography}{10}

\bibitem{Bertone:2004pz}
G.~Bertone, D.~Hooper, and J.~Silk, ``{Particle dark matter: Evidence,
  candidates and constraints},''
  \href{http://dx.doi.org/10.1016/j.physrep.2004.08.031}{{\em Phys. Rept.}
  {\bfseries 405} (2005) 279--390},
\href{http://arxiv.org/abs/hep-ph/0404175}{{\ttfamily arXiv:hep-ph/0404175
  [hep-ph]}}.

\bibitem{Feng:2010gw}
J.~L. Feng, ``{Dark Matter Candidates from Particle Physics and Methods of
  Detection},''
  \href{http://dx.doi.org/10.1146/annurev-astro-082708-101659}{{\em Ann. Rev.
  Astron. Astrophys.} {\bfseries 48} (2010) 495--545},
\href{http://arxiv.org/abs/1003.0904}{{\ttfamily arXiv:1003.0904
  [astro-ph.CO]}}.

\bibitem{Young:2016ala}
B.-L. Young, ``{A survey of dark matter and related topics in cosmology},''
  \href{http://dx.doi.org/10.1007/s11467-017-0680-z,
  10.1007/s11467-016-0583-4}{{\em Front. Phys.(Beijing)} {\bfseries 12} no.~2,
  (2017) 121201}.
[Erratum: Front. Phys.(Beijing)12,no.2,121202(2017)].

\bibitem{Arcadi:2017kky}
G.~Arcadi, M.~Dutra, P.~Ghosh, M.~Lindner, Y.~Mambrini, M.~Pierre, S.~Profumo,
  and F.~S. Queiroz, ``{The waning of the WIMP? A review of models, searches,
  and constraints},''
  \href{http://dx.doi.org/10.1140/epjc/s10052-018-5662-y}{{\em Eur. Phys. J.}
  {\bfseries C78} no.~3, (2018) 203},
\href{http://arxiv.org/abs/1703.07364}{{\ttfamily arXiv:1703.07364 [hep-ph]}}.

\bibitem{Cirelli:2005uq}
M.~Cirelli, N.~Fornengo, and A.~Strumia, ``{Minimal dark matter},''
  \href{http://dx.doi.org/10.1016/j.nuclphysb.2006.07.012}{{\em Nucl. Phys.}
  {\bfseries B753} (2006) 178--194},
\href{http://arxiv.org/abs/hep-ph/0512090}{{\ttfamily arXiv:hep-ph/0512090
  [hep-ph]}}.

\bibitem{Cirelli:2007xd}
M.~Cirelli, A.~Strumia, and M.~Tamburini, ``{Cosmology and Astrophysics of
  Minimal Dark Matter},''
  \href{http://dx.doi.org/10.1016/j.nuclphysb.2007.07.023}{{\em Nucl. Phys.}
  {\bfseries B787} (2007) 152--175},
\href{http://arxiv.org/abs/0706.4071}{{\ttfamily arXiv:0706.4071 [hep-ph]}}.

\bibitem{Cirelli:2008id}
M.~Cirelli, R.~Franceschini, and A.~Strumia, ``{Minimal Dark Matter predictions
  for galactic positrons, anti-protons, photons},''
  \href{http://dx.doi.org/10.1016/j.nuclphysb.2008.03.013}{{\em Nucl. Phys.}
  {\bfseries B800} (2008) 204--220},
\href{http://arxiv.org/abs/0802.3378}{{\ttfamily arXiv:0802.3378 [hep-ph]}}.

\bibitem{Cirelli:2009uv}
M.~Cirelli and A.~Strumia, ``{Minimal Dark Matter: Model and results},''
  \href{http://dx.doi.org/10.1088/1367-2630/11/10/105005}{{\em New J. Phys.}
  {\bfseries 11} (2009) 105005},
\href{http://arxiv.org/abs/0903.3381}{{\ttfamily arXiv:0903.3381 [hep-ph]}}.

\bibitem{Hambye:2009pw}
T.~Hambye, F.~S. Ling, L.~Lopez~Honorez, and J.~Rocher, ``{Scalar Multiplet
  Dark Matter},'' \href{http://dx.doi.org/10.1007/JHEP05(2010)066,
  10.1088/1126-6708/2009/07/090}{{\em JHEP} {\bfseries 07} (2009) 090},
  \href{http://arxiv.org/abs/0903.4010}{{\ttfamily arXiv:0903.4010 [hep-ph]}}.
[Erratum: JHEP05,066(2010)].

\bibitem{Buckley:2009kv}
M.~R. Buckley, L.~Randall, and B.~Shuve, ``{LHC Searches for Non-Chiral Weakly
  Charged Multiplets},'' \href{http://dx.doi.org/10.1007/JHEP05(2011)097}{{\em
  JHEP} {\bfseries 05} (2011) 097},
\href{http://arxiv.org/abs/0909.4549}{{\ttfamily arXiv:0909.4549 [hep-ph]}}.

\bibitem{Cai:2012kt}
Y.~Cai, W.~Chao, and S.~Yang, ``{Scalar Septuplet Dark Matter and Enhanced
  $h\rightarrow \gamma\gamma$ Decay Rate},''
  \href{http://dx.doi.org/10.1007/JHEP12(2012)043}{{\em JHEP} {\bfseries 12}
  (2012) 043},
\href{http://arxiv.org/abs/1208.3949}{{\ttfamily arXiv:1208.3949 [hep-ph]}}.

\bibitem{Earl:2013jsa}
K.~Earl, K.~Hartling, H.~E. Logan, and T.~Pilkington, ``{Constraining models
  with a large scalar multiplet},''
  \href{http://dx.doi.org/10.1103/PhysRevD.88.015002}{{\em Phys. Rev.}
  {\bfseries D88} (2013) 015002},
\href{http://arxiv.org/abs/1303.1244}{{\ttfamily arXiv:1303.1244 [hep-ph]}}.

\bibitem{Cirelli:2014dsa}
M.~Cirelli, F.~Sala, and M.~Taoso, ``{Wino-like Minimal Dark Matter and future
  colliders},'' \href{http://dx.doi.org/10.1007/JHEP10(2014)033,
  10.1007/JHEP01(2015)041}{{\em JHEP} {\bfseries 10} (2014) 033},
  \href{http://arxiv.org/abs/1407.7058}{{\ttfamily arXiv:1407.7058 [hep-ph]}}.
[Erratum: JHEP01,041(2015)].

\bibitem{Ostdiek:2015aga}
B.~Ostdiek, ``{Constraining the minimal dark matter fiveplet with LHC
  searches},'' \href{http://dx.doi.org/10.1103/PhysRevD.92.055008}{{\em Phys.
  Rev.} {\bfseries D92} (2015) 055008},
\href{http://arxiv.org/abs/1506.03445}{{\ttfamily arXiv:1506.03445 [hep-ph]}}.

\bibitem{Cirelli:2015bda}
M.~Cirelli, T.~Hambye, P.~Panci, F.~Sala, and M.~Taoso, ``{Gamma ray tests of
  Minimal Dark Matter},''
  \href{http://dx.doi.org/10.1088/1475-7516/2015/10/026}{{\em JCAP} {\bfseries
  1510} no.~10, (2015) 026},
\href{http://arxiv.org/abs/1507.05519}{{\ttfamily arXiv:1507.05519 [hep-ph]}}.

\bibitem{Garcia-Cely:2015dda}
C.~Garcia-Cely, A.~Ibarra, A.~S. Lamperstorfer, and M.~H.~G. Tytgat,
  ``{Gamma-rays from Heavy Minimal Dark Matter},''
  \href{http://dx.doi.org/10.1088/1475-7516/2015/10/058}{{\em JCAP} {\bfseries
  1510} no.~10, (2015) 058},
\href{http://arxiv.org/abs/1507.05536}{{\ttfamily arXiv:1507.05536 [hep-ph]}}.

\bibitem{Cai:2015kpa}
C.~Cai, Z.-M. Huang, Z.~Kang, Z.-H. Yu, and H.-H. Zhang, ``{Perturbativity
  Limits for Scalar Minimal Dark Matter with Yukawa Interactions: Septuplet},''
  \href{http://dx.doi.org/10.1103/PhysRevD.92.115004}{{\em Phys. Rev.}
  {\bfseries D92} no.~11, (2015) 115004},
\href{http://arxiv.org/abs/1510.01559}{{\ttfamily arXiv:1510.01559 [hep-ph]}}.

\bibitem{DelNobile:2015bqo}
E.~Del~Nobile, M.~Nardecchia, and P.~Panci, ``{Millicharge or Decay: A Critical
  Take on Minimal Dark Matter},''
  \href{http://dx.doi.org/10.1088/1475-7516/2016/04/048}{{\em JCAP} {\bfseries
  1604} no.~04, (2016) 048},
\href{http://arxiv.org/abs/1512.05353}{{\ttfamily arXiv:1512.05353 [hep-ph]}}.

\bibitem{Mahbubani:2005pt}
R.~Mahbubani and L.~Senatore, ``{The Minimal model for dark matter and
  unification},'' \href{http://dx.doi.org/10.1103/PhysRevD.73.043510}{{\em
  Phys. Rev.} {\bfseries D73} (2006) 043510},
\href{http://arxiv.org/abs/hep-ph/0510064}{{\ttfamily arXiv:hep-ph/0510064
  [hep-ph]}}.

\bibitem{DEramo:2007anh}
F.~D'Eramo, ``{Dark matter and Higgs boson physics},''
  \href{http://dx.doi.org/10.1103/PhysRevD.76.083522}{{\em Phys. Rev.}
  {\bfseries D76} (2007) 083522},
\href{http://arxiv.org/abs/0705.4493}{{\ttfamily arXiv:0705.4493 [hep-ph]}}.

\bibitem{Enberg:2007rp}
R.~Enberg, P.~J. Fox, L.~J. Hall, A.~Y. Papaioannou, and M.~Papucci, ``{LHC and
  dark matter signals of improved naturalness},''
  \href{http://dx.doi.org/10.1088/1126-6708/2007/11/014}{{\em JHEP} {\bfseries
  11} (2007) 014},
\href{http://arxiv.org/abs/0706.0918}{{\ttfamily arXiv:0706.0918 [hep-ph]}}.

\bibitem{Cohen:2011ec}
T.~Cohen, J.~Kearney, A.~Pierce, and D.~Tucker-Smith, ``{Singlet-Doublet Dark
  Matter},'' \href{http://dx.doi.org/10.1103/PhysRevD.85.075003}{{\em Phys.
  Rev.} {\bfseries D85} (2012) 075003},
\href{http://arxiv.org/abs/1109.2604}{{\ttfamily arXiv:1109.2604 [hep-ph]}}.

\bibitem{Fischer:2013hwa}
O.~Fischer and J.~J. van~der Bij, ``{The scalar Singlet-Triplet Dark Matter
  Model},'' \href{http://dx.doi.org/10.1088/1475-7516/2014/01/032}{{\em JCAP}
  {\bfseries 1401} (2014) 032},
\href{http://arxiv.org/abs/1311.1077}{{\ttfamily arXiv:1311.1077 [hep-ph]}}.

\bibitem{Cheung:2013dua}
C.~Cheung and D.~Sanford, ``{Simplified Models of Mixed Dark Matter},''
  \href{http://dx.doi.org/10.1088/1475-7516/2014/02/011}{{\em JCAP} {\bfseries
  1402} (2014) 011},
\href{http://arxiv.org/abs/1311.5896}{{\ttfamily arXiv:1311.5896 [hep-ph]}}.

\bibitem{Dedes:2014hga}
A.~Dedes and D.~Karamitros, ``{Doublet-Triplet Fermionic Dark Matter},''
  \href{http://dx.doi.org/10.1103/PhysRevD.89.115002}{{\em Phys. Rev.}
  {\bfseries D89} no.~11, (2014) 115002},
\href{http://arxiv.org/abs/1403.7744}{{\ttfamily arXiv:1403.7744 [hep-ph]}}.

\bibitem{Fedderke:2015txa}
M.~A. Fedderke, T.~Lin, and L.-T. Wang, ``{Probing the fermionic Higgs portal
  at lepton colliders},'' \href{http://dx.doi.org/10.1007/JHEP04(2016)160}{{\em
  JHEP} {\bfseries 04} (2016) 160},
\href{http://arxiv.org/abs/1506.05465}{{\ttfamily arXiv:1506.05465 [hep-ph]}}.

\bibitem{Calibbi:2015nha}
L.~Calibbi, A.~Mariotti, and P.~Tziveloglou, ``{Singlet-Doublet Model: Dark
  matter searches and LHC constraints},''
  \href{http://dx.doi.org/10.1007/JHEP10(2015)116}{{\em JHEP} {\bfseries 10}
  (2015) 116},
\href{http://arxiv.org/abs/1505.03867}{{\ttfamily arXiv:1505.03867 [hep-ph]}}.

\bibitem{Freitas:2015hsa}
A.~Freitas, S.~Westhoff, and J.~Zupan, ``{Integrating in the Higgs Portal to
  Fermion Dark Matter},'' \href{http://dx.doi.org/10.1007/JHEP09(2015)015}{{\em
  JHEP} {\bfseries 09} (2015) 015},
\href{http://arxiv.org/abs/1506.04149}{{\ttfamily arXiv:1506.04149 [hep-ph]}}.

\bibitem{Yaguna:2015mva}
C.~E. Yaguna, ``{Singlet-Doublet Dirac Dark Matter},''
  \href{http://dx.doi.org/10.1103/PhysRevD.92.115002}{{\em Phys. Rev.}
  {\bfseries D92} no.~11, (2015) 115002},
\href{http://arxiv.org/abs/1510.06151}{{\ttfamily arXiv:1510.06151 [hep-ph]}}.

\bibitem{Tait:2016qbg}
T.~M.~P. Tait and Z.-H. Yu, ``{Triplet-Quadruplet Dark Matter},''
  \href{http://dx.doi.org/10.1007/JHEP03(2016)204}{{\em JHEP} {\bfseries 03}
  (2016) 204},
\href{http://arxiv.org/abs/1601.01354}{{\ttfamily arXiv:1601.01354 [hep-ph]}}.

\bibitem{Horiuchi:2016tqw}
S.~Horiuchi, O.~Macias, D.~Restrepo, A.~Rivera, O.~Zapata, and H.~Silverwood,
  ``{The Fermi-LAT gamma-ray excess at the Galactic Center in the
  singlet-doublet fermion dark matter model},''
  \href{http://dx.doi.org/10.1088/1475-7516/2016/03/048}{{\em JCAP} {\bfseries
  1603} no.~03, (2016) 048},
\href{http://arxiv.org/abs/1602.04788}{{\ttfamily arXiv:1602.04788 [hep-ph]}}.

\bibitem{Banerjee:2016hsk}
S.~Banerjee, S.~Matsumoto, K.~Mukaida, and Y.-L.~S. Tsai, ``{WIMP Dark Matter
  in a Well-Tempered Regime: A case study on Singlet-Doublets Fermionic
  WIMP},'' \href{http://dx.doi.org/10.1007/JHEP11(2016)070}{{\em JHEP}
  {\bfseries 11} (2016) 070},
\href{http://arxiv.org/abs/1603.07387}{{\ttfamily arXiv:1603.07387 [hep-ph]}}.

\bibitem{Cai:2016sjz}
C.~Cai, Z.-H. Yu, and H.-H. Zhang, ``{CEPC Precision of Electroweak Oblique
  Parameters and Weakly Interacting Dark Matter: the Fermionic Case},''
  \href{http://dx.doi.org/10.1016/j.nuclphysb.2017.05.015}{{\em Nucl. Phys.}
  {\bfseries B921} (2017) 181--210},
\href{http://arxiv.org/abs/1611.02186}{{\ttfamily arXiv:1611.02186 [hep-ph]}}.

\bibitem{Abe:2017glm}
T.~Abe, ``{Effect of CP violation in the singlet-doublet dark matter model},''
  \href{http://dx.doi.org/10.1016/j.physletb.2017.05.048}{{\em Phys. Lett.}
  {\bfseries B771} (2017) 125--130},
\href{http://arxiv.org/abs/1702.07236}{{\ttfamily arXiv:1702.07236 [hep-ph]}}.

\bibitem{Lu:2016dbc}
W.-B. Lu and P.-H. Gu, ``{Mixed Inert Scalar Triplet Dark Matter, Radiative
  Neutrino Masses and Leptogenesis},''
  \href{http://dx.doi.org/10.1016/j.nuclphysb.2017.09.005}{{\em Nucl. Phys.}
  {\bfseries B924} (2017) 279--311},
\href{http://arxiv.org/abs/1611.02106}{{\ttfamily arXiv:1611.02106 [hep-ph]}}.

\bibitem{Cai:2017wdu}
C.~Cai, Z.-H. Yu, and H.-H. Zhang, ``{CEPC Precision of Electroweak Oblique
  Parameters and Weakly Interacting Dark Matter: the Scalar Case},''
  \href{http://dx.doi.org/10.1016/j.nuclphysb.2017.09.007}{{\em Nucl. Phys.}
  {\bfseries B924} (2017) 128--152},
\href{http://arxiv.org/abs/1705.07921}{{\ttfamily arXiv:1705.07921 [hep-ph]}}.

\bibitem{Maru:2017otg}
N.~Maru, T.~Miyaji, N.~Okada, and S.~Okada, ``{Fermion Dark Matter in
  Gauge-Higgs Unification},''
  \href{http://dx.doi.org/10.1007/JHEP07(2017)048}{{\em JHEP} {\bfseries 07}
  (2017) 048},
\href{http://arxiv.org/abs/1704.04621}{{\ttfamily arXiv:1704.04621 [hep-ph]}}.

\bibitem{Liu:2017gfg}
X.~Liu and L.~Bian, ``{Dark matter and electroweak phase transition in the
  mixed scalar dark matter model},''
  \href{http://dx.doi.org/10.1103/PhysRevD.97.055028}{{\em Phys. Rev.}
  {\bfseries D97} no.~5, (2018) 055028},
\href{http://arxiv.org/abs/1706.06042}{{\ttfamily arXiv:1706.06042 [hep-ph]}}.

\bibitem{Egana-Ugrinovic:2017jib}
D.~Egana-Ugrinovic, ``{The minimal fermionic model of electroweak
  baryogenesis},'' \href{http://dx.doi.org/10.1007/JHEP12(2017)064}{{\em JHEP}
  {\bfseries 12} (2017) 064},
\href{http://arxiv.org/abs/1707.02306}{{\ttfamily arXiv:1707.02306 [hep-ph]}}.

\bibitem{Xiang:2017yfs}
Q.-F. Xiang, X.-J. Bi, P.-F. Yin, and Z.-H. Yu, ``{Exploring Fermionic Dark
  Matter via Higgs Boson Precision Measurements at the Circular Electron
  Positron Collider},''
  \href{http://dx.doi.org/10.1103/PhysRevD.97.055004}{{\em Phys. Rev.}
  {\bfseries D97} no.~5, (2018) 055004},
\href{http://arxiv.org/abs/1707.03094}{{\ttfamily arXiv:1707.03094 [hep-ph]}}.

\bibitem{Voigt:2017vfz}
A.~Voigt and S.~Westhoff, ``{Virtual signatures of dark sectors in Higgs
  couplings},'' \href{http://dx.doi.org/10.1007/JHEP11(2017)009}{{\em JHEP}
  {\bfseries 11} (2017) 009},
\href{http://arxiv.org/abs/1708.01614}{{\ttfamily arXiv:1708.01614 [hep-ph]}}.

\bibitem{Wang:2017sxx}
J.-W. Wang, X.-J. Bi, Q.-F. Xiang, P.-F. Yin, and Z.-H. Yu, ``{Exploring
  triplet-quadruplet fermionic dark matter at the LHC and future colliders},''
  \href{http://dx.doi.org/10.1103/PhysRevD.97.035021}{{\em Phys. Rev.}
  {\bfseries D97} no.~3, (2018) 035021},
\href{http://arxiv.org/abs/1711.05622}{{\ttfamily arXiv:1711.05622 [hep-ph]}}.

\bibitem{DiLuzio:2015oha}
L.~Di~Luzio, R.~Gröber, J.~F. Kamenik, and M.~Nardecchia, ``{Accidental matter
  at the LHC},'' \href{http://dx.doi.org/10.1007/JHEP07(2015)074}{{\em JHEP}
  {\bfseries 07} (2015) 074},
\href{http://arxiv.org/abs/1504.00359}{{\ttfamily arXiv:1504.00359 [hep-ph]}}.

\bibitem{Araki:2011hm}
T.~Araki, C.~Q. Geng, and K.~I. Nagao, ``{Dark Matter in Inert Triplet
  Models},'' \href{http://dx.doi.org/10.1103/PhysRevD.83.075014}{{\em Phys.
  Rev.} {\bfseries D83} (2011) 075014},
\href{http://arxiv.org/abs/1102.4906}{{\ttfamily arXiv:1102.4906 [hep-ph]}}.

\bibitem{Ayazi:2014tha}
S.~Yaser~Ayazi and S.~M. Firouzabadi, ``{Constraining Inert Triplet Dark Matter
  by the LHC and FermiLAT},''
  \href{http://dx.doi.org/10.1088/1475-7516/2014/11/005}{{\em JCAP} {\bfseries
  1411} no.~11, (2014) 005},
\href{http://arxiv.org/abs/1408.0654}{{\ttfamily arXiv:1408.0654 [hep-ph]}}.

\bibitem{Khan:2016sxm}
N.~Khan, ``{Exploring the hyperchargeless Higgs triplet model up to the Planck
  scale},'' \href{http://dx.doi.org/10.1140/epjc/s10052-018-5766-4}{{\em Eur.
  Phys. J.} {\bfseries C78} no.~4, (2018) 341},
\href{http://arxiv.org/abs/1610.03178}{{\ttfamily arXiv:1610.03178 [hep-ph]}}.

\bibitem{Hamada:2015bra}
Y.~Hamada, K.~Kawana, and K.~Tsumura, ``{Landau pole in the Standard Model with
  weakly interacting scalar fields},''
  \href{http://dx.doi.org/10.1016/j.physletb.2015.05.072}{{\em Phys. Lett.}
  {\bfseries B747} (2015) 238--244},
\href{http://arxiv.org/abs/1505.01721}{{\ttfamily arXiv:1505.01721 [hep-ph]}}.

\bibitem{Foot:1988aq}
R.~Foot, H.~Lew, X.~G. He, and G.~C. Joshi, ``{Seesaw Neutrino Masses Induced
  by a Triplet of Leptons},''
\href{http://dx.doi.org/10.1007/BF01415558}{{\em Z. Phys.} {\bfseries C44}
  (1989) 441}.

\bibitem{Minkowski:1977sc}
P.~Minkowski, ``{$\mu \to e\gamma$ at a Rate of One Out of $10^{9}$ Muon
  Decays?},''
\href{http://dx.doi.org/10.1016/0370-2693(77)90435-X}{{\em Phys. Lett.}
  {\bfseries 67B} (1977) 421--428}.

\bibitem{GellMann:1980vs}
M.~Gell-Mann, P.~Ramond, and R.~Slansky, ``{Complex Spinors and Unified
  Theories},'' {\em Conf. Proc.} {\bfseries C790927} (1979) 315--321,
\href{http://arxiv.org/abs/1306.4669}{{\ttfamily arXiv:1306.4669 [hep-th]}}.

\bibitem{Yanagida:1979as}
T.~Yanagida, ``{HORIZONTAL SYMMETRY AND MASSES OF NEUTRINOS},''
{\em Conf. Proc.} {\bfseries C7902131} (1979) 95--99.

\bibitem{Mohapatra:1979ia}
R.~N. Mohapatra and G.~Senjanovic, ``{Neutrino Mass and Spontaneous Parity
  Violation},''
\href{http://dx.doi.org/10.1103/PhysRevLett.44.912}{{\em Phys. Rev. Lett.}
  {\bfseries 44} (1980) 912}.

\bibitem{Kannike:2012pe}
K.~Kannike, ``{Vacuum Stability Conditions From Copositivity Criteria},''
  \href{http://dx.doi.org/10.1140/epjc/s10052-012-2093-z}{{\em Eur. Phys. J.}
  {\bfseries C72} (2012) 2093},
\href{http://arxiv.org/abs/1205.3781}{{\ttfamily arXiv:1205.3781 [hep-ph]}}.

\bibitem{Griest:1990kh}
K.~Griest and D.~Seckel, ``{Three exceptions in the calculation of relic
  abundances},''
\href{http://dx.doi.org/10.1103/PhysRevD.43.3191}{{\em Phys. Rev.} {\bfseries
  D43} (1991) 3191--3203}.

\bibitem{Ade:2015xua}
{\bfseries Planck} Collaboration, P.~A.~R. Ade {\em et~al.}, ``{Planck 2015
  results. XIII. Cosmological parameters},''
  \href{http://dx.doi.org/10.1051/0004-6361/201525830}{{\em Astron. Astrophys.}
  {\bfseries 594} (2016) A13},
\href{http://arxiv.org/abs/1502.01589}{{\ttfamily arXiv:1502.01589
  [astro-ph.CO]}}.

\bibitem{Ahnen:2016qkx}
{\bfseries Fermi-LAT, MAGIC} Collaboration, M.~L. Ahnen {\em et~al.}, ``{Limits
  to dark matter annihilation cross-section from a combined analysis of MAGIC
  and Fermi-LAT observations of dwarf satellite galaxies},''
  \href{http://dx.doi.org/10.1088/1475-7516/2016/02/039}{{\em JCAP} {\bfseries
  1602} no.~02, (2016) 039},
\href{http://arxiv.org/abs/1601.06590}{{\ttfamily arXiv:1601.06590
  [astro-ph.HE]}}.

\bibitem{Mitridate:2017izz}
A.~Mitridate, M.~Redi, J.~Smirnov, and A.~Strumia, ``{Cosmological Implications
  of Dark Matter Bound States},''
  \href{http://dx.doi.org/10.1088/1475-7516/2017/05/006}{{\em JCAP} {\bfseries
  1705} no.~05, (2017) 006},
\href{http://arxiv.org/abs/1702.01141}{{\ttfamily arXiv:1702.01141 [hep-ph]}}.

\bibitem{Hisano:2015rsa}
J.~Hisano, K.~Ishiwata, and N.~Nagata, ``{QCD Effects on Direct Detection of
  Wino Dark Matter},'' \href{http://dx.doi.org/10.1007/JHEP06(2015)097}{{\em
  JHEP} {\bfseries 06} (2015) 097},
\href{http://arxiv.org/abs/1504.00915}{{\ttfamily arXiv:1504.00915 [hep-ph]}}.

\bibitem{Cui:2017nnn}
{\bfseries PandaX-II} Collaboration, X.~Cui {\em et~al.}, ``{Dark Matter
  Results From 54-Ton-Day Exposure of PandaX-II Experiment},''
  \href{http://dx.doi.org/10.1103/PhysRevLett.119.181302}{{\em Phys. Rev.
  Lett.} {\bfseries 119} no.~18, (2017) 181302},
\href{http://arxiv.org/abs/1708.06917}{{\ttfamily arXiv:1708.06917
  [astro-ph.CO]}}.

\bibitem{Aprile:2018dbl}
{\bfseries XENON} Collaboration, E.~Aprile {\em et~al.}, ``{Dark Matter Search
  Results from a One Ton-Year Exposure of XENON1T},''
  \href{http://dx.doi.org/10.1103/PhysRevLett.121.111302}{{\em Phys. Rev.
  Lett.} {\bfseries 121} no.~11, (2018) 111302},
\href{http://arxiv.org/abs/1805.12562}{{\ttfamily arXiv:1805.12562
  [astro-ph.CO]}}.

\bibitem{Adler:1969gk}
S.~L. Adler, ``{Axial vector vertex in spinor electrodynamics},''
\href{http://dx.doi.org/10.1103/PhysRev.177.2426}{{\em Phys. Rev.} {\bfseries
  177} (1969) 2426--2438}.

\bibitem{Bell:1969ts}
J.~S. Bell and R.~Jackiw, ``{A PCAC puzzle: $\pi^0\to\gamma\gamma$ in the
  $\sigma$-model},''
\href{http://dx.doi.org/10.1007/BF02823296}{{\em Nuovo Cim.} {\bfseries A60}
  (1969) 47--61}.

\bibitem{Witten:1982fp}
E.~Witten, ``{An SU(2) Anomaly},''
\href{http://dx.doi.org/10.1016/0370-2693(82)90728-6}{{\em Phys. Lett.}
  {\bfseries 117B} (1982) 324--328}.

\bibitem{Bar:2002sa}
O.~Bar, ``{On Witten's global anomaly for higher SU(2) representations},''
  \href{http://dx.doi.org/10.1016/S0550-3213(02)01027-1}{{\em Nucl. Phys.}
  {\bfseries B650} (2003) 522--542},
\href{http://arxiv.org/abs/hep-lat/0209098}{{\ttfamily arXiv:hep-lat/0209098
  [hep-lat]}}.

\bibitem{Patrignani:2016xqp}
{\bfseries Particle Data Group} Collaboration, C.~Patrignani {\em et~al.},
  ``{Review of Particle Physics},''
\href{http://dx.doi.org/10.1088/1674-1137/40/10/100001}{{\em Chin. Phys.}
  {\bfseries C40} no.~10, (2016) 100001}.

\bibitem{Beringer:1900zz}
{\bfseries Particle Data Group} Collaboration, J.~Beringer {\em et~al.},
  ``{Review of Particle Physics (RPP)},''
\href{http://dx.doi.org/10.1103/PhysRevD.86.010001}{{\em Phys. Rev.} {\bfseries
  D86} (2012) 010001}.

\bibitem{Hambye:1996wb}
T.~Hambye and K.~Riesselmann, ``{Matching conditions and Higgs mass upper
  bounds revisited},'' \href{http://dx.doi.org/10.1103/PhysRevD.55.7255}{{\em
  Phys. Rev.} {\bfseries D55} (1997) 7255--7262},
\href{http://arxiv.org/abs/hep-ph/9610272}{{\ttfamily arXiv:hep-ph/9610272
  [hep-ph]}}.

\end{thebibliography}\endgroup

\end{document}